\documentclass[%
reprint,
superscriptaddress,
 amsmath,amssymb,aps,
pre,
]{revtex4-2}

\usepackage{graphicx}
\usepackage{bm}
\usepackage{tikz}
\usepackage{hyperref}

\begin{document}


\title{
Phase diffusion and fluctuations in a dissipative Bose-Josephson junction 
}

\author{Abhik Kumar Saha}
\affiliation{School of Physical Sciences, Indian Association for the Cultivation of Science,
Jadavpur, Kolkata 700032, India.}
\author{Deb Shankar Ray}%
\email{pcdsr@iacs.res.in}
\affiliation{School of Chemical Sciences, Indian Association for the Cultivation of Science,
Jadavpur, Kolkata 700032, India.}

\author{Bimalendu Deb}
\email{Corresponding author: msbd@iacs.res.in}
\affiliation{School of Physical Sciences, Indian Association for the Cultivation of Science,
Jadavpur, Kolkata 700032, India.}

\date{\today}

\begin{abstract}
We analyze the phase diffusion, quantum fluctuations and their spectral features of an one-dimensional Bose-Josephson junction (BJJ) coupled to a bosonic heat bath. We show the dependence of the phase diffusion coefficient on the on-site interaction parameter $U$ and the temperature in zero-phase and $\pi$-phase modes. We find that in the $\pi$-phase mode, the phase diffusion co-efficient as a function of $U$ decreases so long as $U$ is below a critical value while it increases above the critical value. This criticality of on-site interaction reflects a transition between Josephson oscillation and  macroscopic quantum self-trapping (MQST) regime. Based on the thermal canonical Wigner distribution, we calculate the coherence factor to understand its dependence on temperature and on-site interaction energy in Josephson oscillation and MQST regime. Furthermore, we discuss coherent and incoherent spectral properties in connection with the fluctuations of the relative phase and the population imbalance in both zero and $\pi$-phase modes from weak to strong dissipation regime.

\end{abstract}

\maketitle
\clearpage
\section{Introduction}
A dynamical system in contact with a reservoir has been a subject of wide attention in dissipative dynamics \cite{ch6_1,ch6_2}. Over the years, the dissipative quantum systems under the effect of random noise have been extensively investigated both theoretically and experimentally \cite{ch6_3,ch6_4,ch6_5,ch6_6,ch6_7} in widely different areas such as condensed matter physics, quantum optics, magnetic resonance spectroscopy etc.
In recent years, the study of quantum dissipation in ultracold atomic gases has attracted much attention due to the presence of various loss processes affecting the coherence of the atomic matter waves \cite{ch6_8,ch6_9,ch6_10}. Also, an ultracold atomic system has become a test bed to study dissipation in an out-of-equilibrium quantum many-body system  \cite{ch6_11,ch6_12,ch6_13,ch6_14,ch6_15,ch6_15a}. Apart from dephasing and relaxation dynamics, the combined effect of interaction and dissipation in open quantum systems can give rise to non-equilibrium steady states and transitions between them \cite{ch6_16,ch6_17,ch6_18,ch6_19}. The BJJ is an ideal system where both the effects of interaction and dissipation can be explored to understand coherent and incoherent quantum dynamics of matter waves \cite{ch6_20,ch6_20a}.

There are various ways in which the dissipation can be introduced in the system. The presence of intrinsic coupling between the Josephson mode and the quasi-particle mode often termed as sound mode leads to damping of oscillation of population imbalance in a one-dimensional (1D) BJJ \cite{ch6_21,ch6_22}. Moreover, impurity models can be realized in an ultracold atomic system by coupling the impurity atom with the phonon bath of a condensate \cite{ch6_23,ch6_24,ch6_25,ch6_26}, resulting in dissipation in the system. Dissipation also originates from the finite temperature effects \cite{ch6_27,ch6_28,ch6_29}, and the coupling of the system with the external environment \cite{ch6_30}. In this context, phase diffusion plays a fundamental role in the dynamical behaviour of \cite{ch6_31,ch6_32,ch6_33,ch6_34} cold atoms in optical lattices and Bose-Einstein condensates (BECs) \cite{ch6_35,ch6_36,ch6_37,ch6_38}. There are many theoretical and experimental works suggesting that the interaction between the particles can lead to phase diffusion \cite{ch6_38a,ch6_38b,ch6_38c,ch6_38d,ch6_38e,ch6_38f,ch6_38g,ch6_31,ch6_32,ch6_33,ch6_34}. Phase diffusion and heating effect in a BJJ due to thermal fluctuations have already been observed in experiment \cite{ch6_39}. Motivated by these recent advances, we examine the phase diffusion, the coherent and incoherent phenomena in a 1D dissipative BJJ. The question we address here is how the phase diffusion coefficient depends on the interaction parameter and temperature in both zero and $\pi$-phase modes of 1D dissipative BJJ.

In this paper, we consider a two-mode Hamiltonian of BJJ coupled to bosonic heat baths. The nonlinear interaction between the system and an external bosonic bath induces the dissipation and the phase diffusion. This is distinct from the treatment of the system having intrinsic coupling between Josephson mode and the quasi-particle mode. The aim of the present work is to derive analytically the dissipative BJJ equations in presence of quantum noise to probe the Langevin dynamics of relative phase. We derive the analytical formula for the phase diffusion coefficient for zero and $\pi$-phase modes and observe that the phase diffusion coefficient depends on the on-site interaction and temperature of the system. In the zero-phase mode, it becomes comparatively large for small dissipation in the high temperature regime. On the other hand, in the $\pi$-phase mode it exhibits an interesting phase transition-like behaviour between the Josephson oscillation and the MQST regime. We observe that in the $\pi$-phase mode, the phase diffusion co-efficient decreases so long as the on-site interaction is below a critical value while it increases above the critical value. Making use of the thermal canonical Wigner distribution \cite{ch6_40}, we calculate the coherence factor \cite{ch6_41} to capture its dependence on interaction energy and temperature. Our analysis of coherence factor reveals that the 1D dissipative BJJ has a higher degree of coherence which is in good agreement with the experimental observation \cite{ch6_39}. The underlying quantum fluctuations in population imbalance and relative phase have been analyzed in terms of fluctuation spectra to demonstrate the coherent and incoherent behaviour of the system from weak to strong dissipation limit in both zero and $\pi$-phase modes.

The paper is organized in the following way. In section \ref{sec:2}, we derive the dissipative BJJ equations by coupling the two-mode Hamiltonian to the bosonic baths in presence of noise to study the quantum and thermal properties of noise. In section \ref{sec:3}, we analyze theoretically phase diffusion coefficient in 1D dissipative BJJ. In section \ref{sec:4}, we derive the analytical formula for the fluctuation spectrum of population imbalance and phase difference. In section \ref{sec:5}, we present and discuss our results on numerical simulations of coherence factor, phase diffusion coefficient and the spectra of the fluctuation to corroborate the theoretical scheme. The paper is concluded in section \ref{sec:6}. 
\section{Dissipative Bose-Josephson junction: A phase diffusion model}\label{sec:2}
The Gross-Pitaevskii Hamiltonian for a system of $N$ bosons at zero temperature is given by 
\begin{eqnarray}
\hat H_{\rm GP}=\hat H_{0}+ \hat H_{\rm int}
\label{1}
\end{eqnarray}
where
\begin{eqnarray}
\hat H_{0}=\int d{\bf r}\left[-\frac{\hbar^2}{2m}\hat \psi^{\dagger}\nabla^{2}\hat \psi+\hat\psi^{\dagger}V_{\rm ext}\hat\psi\right] 
\end{eqnarray}
\begin{eqnarray}
\hat H_{\rm int}=\frac{2\pi\hbar^{2}a_{s}}{m}\int d{\bf r} \hat\psi^{\dagger}\hat\psi^{\dagger}\hat\psi\hat\psi  
\end{eqnarray}
$\hat\psi$ and $\hat\psi^{\dagger}$ represent bosonic fields and $V_{\rm ext}$ is the external trap potential \cite{ch6_41a,ch6_41b} of the form $V_{\rm ext}=V(\rho)+V_{\rm dw}(x)$, where $V(\rho)=\frac{1}{2}m\omega_{\rho}^2\rho^2$ which permits harmonic oscillations with frequency $\omega_{\rho}$ along radial directions i.e, $y$- and $\textrm{z}$- axes and a symmetric double-well (DW) potential $V_{\rm dw}(x)$ along $x$- axis. Here $\rho^2=y^2+\textrm{z}^2$, $a_{s}$ denotes the $s$-wave scattering length and $m$ is the atomic mass.
In the strong radial confinement regime ($\omega_{\rho}\gg\omega_{x}$) where $\omega_{x}$ be the axial frequency, we assume that all the atoms occupy the ground state of the radial harmonic potential. To proceed further, we integrate over the radial harmonic oscillator states and obtain an effective 1D Hamiltonian for the system. The lowest two energy eigen-functions are quasi-degenerate. For symmetric DW, the lowest eigenstate $\phi_{\rm g}$  is space-symmetric $(\phi_{\rm g}(x)=\phi_{\rm g}(-x))$ and the other quasi-degenerate state $\phi_{\rm e}(x)$ is anti-symmetric $(\phi_{\rm e}(x)=-\phi_{\rm e}(-x))$.

The wave function $\hat \psi$ can be written as
\begin{eqnarray}
\hat\psi=\hat a_{g}\phi_{g}+\hat a_{e}\phi_{e}  
\end{eqnarray}
with $\hat a_{g}$ and $\hat a_{e}$ ($\hat a^{\dagger}_{g}$ and $\hat a^{\dagger}_{e}$ ) being the annihilation (creation) operators for a particle in the ground and first excited states, respectively. The operators obey the standard bosonic commutation relation $[\hat a_{i},\hat a^{\dagger}_{j}]=\delta_{ij}$. By defining further two operators $\hat a_{L}=\frac{1}{\sqrt{2}}(\hat a_{g}+\hat a_{e})$ and $\hat a_{R}=\frac{1}{\sqrt{2}}(\hat a_{g}-\hat a_{e})$ and their hermitian counterparts, the effective 1D wave-function becomes
\begin{eqnarray}
\hat\psi=\hat a_{L}\phi_{+}+\hat a_{R}\phi_{-} 
\end{eqnarray}
where, $\phi_{+}=\frac{1}{\sqrt{2}}(\phi_{g}+\phi_{e})$ and $\phi_{-}=\frac{1}{\sqrt{2}}(\phi_{g}-\phi_{e})$.
Making use of this wave function in the Gross-Pitaevskii Hamiltonian of Eq. (\ref{1}), we obtain a two-mode Hamiltonian of the form
\begin{eqnarray}
\hat H_{\rm TM} &=& \hat a^{\dagger}_{L}\hat a_{L}E_{1}+\hat a^{\dagger}_{R}\hat a_{R}E_{2}-(\hat a^{\dagger}_{L}\hat a_{R}+\hat a^{\dagger}_{R}\hat a_{L})K  \nonumber\\ &+&\frac{U_{+}}{2}\hat a^{\dagger}_{L}\hat a^{\dagger}_{L}\hat a_{L}\hat a_{L}+\frac{U_{-}}{2}\hat a^{\dagger}_{R}\hat a^{\dagger}_{R}\hat a_{R}\hat a_{R} 
\end{eqnarray}
where
\begin{eqnarray}
K=-\int\left[\frac{\hbar^2}{2m}(\nabla_{x}\phi_{+}\nabla_{x}\phi_{-}) +\phi_{+}V_{\rm dw}(x)\phi_{-}\right]dx\nonumber
\end{eqnarray}
\begin{eqnarray}
 E_{1(2)}=\int\left[\frac{\hbar^2}{2m}|\nabla_{x}\phi_{+(-)}|^2+|\phi_{+(-)}|^2V_{dw}(x)\right]dx\nonumber
\end{eqnarray}
and
\begin{eqnarray}
 U_{+(-)}=\frac{4\pi\hbar^2a_{s}}{m}\int|\phi_{+(-)}|^{4}dx\nonumber
\end{eqnarray}
where, $\nabla_{x}\equiv\frac{\partial}{\partial x}$, $K$ is the tunneling amplitude between two sites of the DW and $U_{+(-)}$ is the on-site interaction strength for left(right) side of the DW arising out of nonlinearity. For a symmetric DW potential, we write $E_{1}=E_{2}=E$. 

\subsection{Two-mode model coupled to bosonic heat baths}
Usually, Josephson oscillations in a DW potential is nondissipative meaning that the dynamics of the atom number imbalance and relative phase remains undamped over time \cite{ch6_41c,ch6_41d,ch6_41e}. However, in recent times several studies reported a dissipative BJJ which is analogous to a pendulum with friction \cite{ch6_41f,ch6_41g,ch6_41h,ch6_46a}. In order to study the effects of dissipation in BJJ, we consider a model of a BJJ coupled to two bosonic baths described by the total Hamiltonian
\begin{eqnarray}
\hat H_{\rm T}&=&\hat H_{\rm TM}+\hbar\sum_{k, m=L,R} \omega_{k}(\hat b^{\dagger}_{\rm mk}\hat b_{\rm mk})\nonumber\\
&+&\hbar\sum_{k, m=L,R}g_{k}\hat a^{\dagger}_{m}\hat a_{m}(\hat b^{\dagger}_{\rm mk}+\hat b_{\rm mk})
\label{eqn:7}
\end{eqnarray}
where $\hat b_{mk}$ and $\hat b^{\dagger}_{mk}$ are the bosonic annihilation and creation operators respectively, corresponding to $k$-th bath mode and $m$-th well. $\omega_{k}$ represents the frequency of the $k$-th bath mode. $g_{k}$ represents the coupling between the $k$-th bath mode and the on-site boson number. The bath-BJJ coupling constants for the $k$-th bath mode for both the wells are assumed to be same. It is important to emphasize that the coupling between the system and the bath modes are nonlinear \cite{ch6_30}. Also the excitation in the bath modes is not accompanied by energetic deexcitation of the system as in the usual linear system-bath coupling \cite{ch6_41b}. The influence of bath and the non-linear coupling of the form (\ref{eqn:7}) have been used earlier \cite{ch6_30} and this gives rise to dissipation and the phase diffusion of the system. The Heisenberg equations of motion for the system and the bath operators are given by the following equations
\begin{eqnarray}
\dot{\hat {a}}_{L} &=&-\frac{iE}{\hbar}\hat a_{L}+\frac{iK}{\hbar}\hat a_{R}-\frac{iU_{+}}{\hbar}\hat a^{\dagger}_{L}\hat a_{L}\hat a_{L}-i\sum_{k} g_{k}\hat b^{\dagger}_{Lk}(t)a_{L}\nonumber\\
&-& i\sum_{k} g_{k}\hat b_{Lk}(t)a_{L}
\label{eqn:8}
\end{eqnarray}
\begin{eqnarray}
\dot{\hat {a}}_{R}&=&-\frac{iE}{\hbar}\hat a_{R}+\frac{iK}{\hbar}\hat a_{L}-\frac{iU_{-}}{\hbar}\hat a^{\dagger}_{R}\hat a_{R}\hat a_{R}-i\sum_{k} g_{k}\hat b^{\dagger}_{Rk}(t)a_{R}\nonumber\\
&-& i\sum_{k} g_{k}\hat b_{Rk}(t)a_{R}
\label{eqn:9}
\end{eqnarray}
\begin{eqnarray}
\dot{\hat {b}}_{Lk}=-i\omega_{k}\hat b_{Lk}-ig_{k}\hat a^{\dagger}_{L}\hat a_{L}(t) 
\label{eqn:10}
\end{eqnarray}
\begin{eqnarray}
\dot{\hat {b}}_{Rk}=-i\omega_{k}\hat b_{Rk}-ig_{k}\hat a^{\dagger}_{R}\hat a_{R}(t)
\label{eqn:11}
\end{eqnarray}
Formal integration of Eqs. (\ref{eqn:10}) and (\ref{eqn:11}) yields
\begin{eqnarray}
\hat b_{Lk}(t)&=&\hat b_{Lk}(t_{0})e^{-i\omega_{k}(t-t_{0})}\nonumber\\
&-&ig_{k}\int_{t_{0}}^{t}\hat a^{\dagger}_{L}(t') \hat a_{L}(t')e^{-i\omega_{k}(t-t')}dt'
\label{eqn:12}
\end{eqnarray}
\begin{eqnarray}
\hat b_{Rk}(t)&=&\hat b_{Rk}(t_{0})e^{-i\omega_{k}(t-t_{0})}\nonumber\\
&-&ig_{k}\int_{t_{0}}^{t}\hat a^{\dagger}_{R}(t') \hat a_{R}(t')e^{-i\omega_{k}(t-t')}dt'
\label{eqn:13}
\end{eqnarray}
where the first term is the free evolution of bath operators whereas the second term arises due to the interaction with the system. Inserting Eqs. (\ref{eqn:12}) and (\ref{eqn:13}) in Eqs. (\ref{eqn:8}) and (\ref{eqn:9}) we get 
\begin{eqnarray}
\dot{\hat {a}}_{L}&=&-\frac{iE}{\hbar}\hat a_{L}+\frac{iK}{\hbar}\hat a_{R}-\frac{iU_{+}}{\hbar}\hat a^{\dagger}_{L}\hat a_{L}\hat a_{L}\nonumber\\
&-& i\sum_{k} g_{k}\hat b_{Lk}(t_{0})e^{-i\omega_{k}(t-t_{0})}\hat a_{L}\nonumber\\
&-&\sum_{k}g^{2}_{k}\int_{t_{0}}^{t}\hat a^{\dagger}_{L}(t') \hat a_{L}(t') \hat a_{L}(t) e^{-i\omega_{k}(t-t')}dt' 
\label{eqn:14a}
\end{eqnarray}
\begin{eqnarray}
\dot{\hat {a}}_{R}&=&-\frac{iE}{\hbar}\hat a_{R}+\frac{iK}{\hbar}\hat a_{L}-\frac{iU_{-}}{\hbar}\hat a^{\dagger}_{R}\hat a_{R}\hat a_{R}\nonumber\\
&-& i\sum_{k} g_{k}\hat b_{Rk}(t_{0})e^{-i\omega_{k}(t-t_{0})}\hat a_{R}\nonumber\\
&-&\sum_{k}g^{2}_{k}\int_{t_{0}}^{t}\hat a^{\dagger}_{R}(t') \hat a_{R}(t') \hat a_{R}(t) e^{-i\omega_{k}(t-t')}dt'
\label{eqn:15a}
\end{eqnarray}
Changing the integration variable from $t'$ to $\tau=t-t'$ in Eq. (\ref{eqn:14a}), we may write the last term in Eq. (\ref{eqn:14a}) as $\sum_{k}g^{2}_{k}\int_{0}^{t-t_{0}}\hat a^{\dagger}_{L}(t-\tau) \hat a_{L}(t-\tau) \hat a_{L}(t-\tau) e^{-i\omega_{k}\tau}d\tau$, where we have approximated $a_{L}(t)=a_{L}(t-\tau)$ for the last annihilation operator, since the interference time $\tau_{c}$ of $\sum_{k}g^{2}_{k}e^{-i\omega_{k}\tau}$ is much smaller than the time over which the amplitude and the phase modulation of $a_{L}(t)$ take place. Thus for times $t-t_{0}>\tau_{c}$, $\tau_{c}\rightarrow 0$, the summation acts as a delta function so that we may write the integral approximately as $\hat a^{\dagger}_{L} \hat a_{L} \hat a_{L}\int_{0}^{\infty}d\tau \sum_{k}g^{2}_{k} e^{-i(E/\hbar-\omega_{k})\tau}$. Assuming that the bath modes are closely spaced in frequency we replace the summation over $k$ by an integral over $\omega$, i.e $\sum_{k}\rightarrow \int d\omega \Theta (\omega)$, where $\Theta(\omega)$ is the density of states. This density and $g(\omega)$ are proportional to the powers of $\omega$ and vary very little in the frequency interval $\tau^{-1}$ over $\omega$. This leads us to following two equations for the reduced dynamics (same procedure is followed for Eq. (\ref{eqn:15a}))
\begin{eqnarray}
\dot{\hat {a}}_{L} &=& -\frac{iE}{\hbar}\hat a_{L}+\frac{iK}{\hbar}\hat a_{R}-\frac{iU_{+}}{\hbar}\hat a^{\dagger}_{L}\hat a_{L}\hat a_{L}
-\frac{\gamma}{2}\hat a^{\dagger}_{L}\hat a_{L} \hat a_{L}\nonumber\\
&+&\hat f_{L}(t)\hat a_{L} 
\end{eqnarray}
\begin{eqnarray}
\dot{\hat {a}}_{R} &=& -\frac{iE}{\hbar}\hat a_{R}+\frac{iK}{\hbar}\hat a_{L}-\frac{iU_{-}}{\hbar}\hat a^{\dagger}_{R}\hat a_{R}\hat a_{R}-\frac{\gamma}{2}\hat a^{\dagger}_{R}\hat a_{R} \hat a_{R}\nonumber\\
&+&\hat f_{R}(t)\hat a_{R} 
\end{eqnarray}
where, $\gamma=2\pi g^{2}(\Omega)\Theta(\Omega)$ represents the dissipation of the modes and $\Omega=E/\hbar$. $\Theta(\Omega)$ is the density of the bath modes. The terms $\hat f_{L}(t)=-i\sum_{k} g_{k}(t_{0})\hat b_{Lk}(t_{0})e^{-i\omega_{k}(t-t_{0})}$ and  $\hat f_{R}(t)=-i\sum_{k} g_{k}(t_{0})\hat b_{Rk}(t_{0})e^{-i\omega_{k}(t-t_{0})}$ refer to quantum noise due to the heat baths for the $L$ and $R$ modes. Now, eliminating the high frequency oscillation term using the transformation $\hat A_{L,R}=\hat a_{L,R}e^{i\Omega(t-t_{0})} $,  the above equations becomes
\begin{eqnarray}
\dot{\hat {A}}_{L} &=& \frac{iK}{\hbar}\hat A_{R}-\frac{iU_{+}}{\hbar}\hat A^{\dagger}_{L}\hat A_{L}\hat A_{L}-\frac{\gamma}{2}\hat A^{\dagger}_{L}\hat A_{L} \hat A_{L}\nonumber\\
&+&\hat F_{L}(t)\hat a_{L} 
\label{eqn:18}
\end{eqnarray}
\begin{eqnarray}
\dot{\hat {A}}_{R} &=& \frac{iK}{\hbar}\hat A_{L}-\frac{iU_{-}}{\hbar}\hat A^{\dagger}_{R}\hat A_{R}\hat A_{R}-\frac{\gamma}{2}\hat A^{\dagger}_{R}\hat A_{R} \hat A_{R}\nonumber\\
&+&\hat F_{R}(t)\hat a_{R}
\label{eqn:19}
\end{eqnarray}
where ,
\begin{eqnarray}
\hat F_{L}(t)=-i\sum_{k} g_{k}(t_{0})\hat b_{Lk}(t_{0})e^{-i(\omega_{k}-\Omega)(t-t_{0})}\label{eqn:20}\\ 
\hat F_{R}(t)=-i\sum_{k} g_{k}(t_{0})\hat b_{Rk}(t_{0})e^{-i(\omega_{k}-\Omega)(t-t_{0})} \label{eqn:21}\\\nonumber
\end{eqnarray}
refer to the quantum noise due to the heat baths modulated by oscillation of the system for the $L$ and $R$ modes. Eqs.  (\ref{eqn:18}) and (\ref{eqn:19}) and the noise operators in Eqs. (\ref{eqn:20}) and (\ref{eqn:21}) appear as a natural consequence of system-reservoir coupling, Born-Markov and secular approximation \cite{ch6_42,ch6_43,ch6_44}. 

To construct quantum Langevin equation with $c$-number noise \cite{ch6_45,ch6_46}, we return to Eq. (\ref{eqn:18}) and (\ref{eqn:19}) and carry out quantum mechanical average over the initial product separable quantum states of the system oscillator and the bath oscillator at $t_{0}=0$ $|\alpha\rangle |\mu_{1}\rangle |\mu_{2}\rangle .... |\mu_{k}\rangle...|\mu_{N}\rangle$. Here $|\alpha\rangle$ refers to the initial coherent state of the system and $\{|\mu_{k}\rangle\}$ corresponds to the initial coherent states of the bath operators. We denote the quantum mechanical averages for the system and the bath operators as $\langle \hat A_{L,R} \rangle= \alpha_{L,R}$, $\langle \hat A^{\dagger}_{L,R} \rangle= \alpha^{*}_{L,R}$, and $\langle \hat F_{L,R} \rangle= \xi_{L,R}$. Here $\xi_{L,R}=-i\sum_{k} g_{k}(t_{0})\mu_{Lk,Rk}e^{-i(\omega_{k}-\Omega)(t-t_{0})}$. The $c$-number amplitudes may now be written as  $\alpha_{L}=\sqrt{N_{L}}e^{i\theta_{L}}$ and $\alpha_{R}=\sqrt{N_{R}}e^{i\theta_{R}}$. $N_{L(R)}$ is the number of atoms in the left (right) well. $\theta_{L(R)}$ is the phase of the atoms in the left (right) well. The complex amplitude equations for two wells are given by
\begin{eqnarray}
\dot \alpha_{L}=\frac{iK}{\hbar}\alpha_{R}-\frac{iU_{+}}{\hbar}N_{L}\alpha_{L}-\frac{\gamma}{2}N_{L}\alpha_{L}+\xi_{L}(t)\alpha_{L} 
\end{eqnarray}
\begin{eqnarray}
\dot \alpha_{R}=\frac{iK}{\hbar}\alpha_{L}-\frac{iU_{-}}{\hbar}N_{R}\alpha_{R}-\frac{\gamma}{2}N_{R}\alpha_{R}+\xi_{R}(t)\alpha_{R} 
\end{eqnarray} 
Separating the real and imaginary parts of the above equations we obtain after some algebra the following equations in terms of the normalized atom number imbalance $z(t)$ and phase difference $\theta(t)$ as follows: 
\begin{widetext}
\begin{eqnarray}
\dot{z}(t)=-\frac{2K}{\hbar}{\sqrt{1-z^2(t)}}\sin\theta(t)-\gamma N z(t)+\xi_{z}
\label{eqn:imbalance}
\end{eqnarray}
\begin{eqnarray}
\dot{\theta}(t)&=&\frac{2K}{\hbar}\left[\frac{z(t)}{\sqrt{1-z^2(t)}}\cos\theta(t)+\frac{U_{+}-U_{-}}{4K}N
+\frac{UN}{2K}z(t)\right]+ \xi_{\theta}     
\end{eqnarray}
\end{widetext}
where, $\xi_{z}=2 \xi_{L}(t)N_{L}/N-2 \xi_{R}(t)N_{R}/N$, $\xi_{\theta}=\xi_{R}-\xi_{L}$ and
\begin{eqnarray}
z=\frac{|\alpha_{L}|^2-|\alpha_{R}|^2}{|\alpha_{L}|^2+|\alpha_{R}|^2}=\frac{N_{L}-N_{R}}{N_{L}+N_{R}} 
\end{eqnarray}
The conjugate variable is relative phase defined by
\begin{eqnarray}
\theta=\theta_{R}-\theta_{L} 
\end{eqnarray}
where, $U=\frac{U_{+}+U_{-}}{2}$. Now for symmetric DW, we have $U_{+}=U_{-}=U$ and the phase equation reduces to 
\begin{eqnarray}
\dot{\theta}(t)=\frac{2K}{\hbar}\left[\frac{z(t)}{\sqrt{1-z^2(t)}}\cos\theta(t)+\Lambda_{0} z(t)\right]+\xi_{\theta}
\label{eqn:phase}
\end{eqnarray}
where, $\Lambda_{0}=\frac{NU}{2K}$ characterizes the many-body interaction parameter with $U$ being the on-site mean two-body interaction energy. Equations (\ref{eqn:imbalance}) and (\ref{eqn:phase}) represent the dissipative BJJ equation with noise. In the absence of noise ($\xi_{L,R}(t)=0$) the equations reduce to the dissipative BJJ equations in which the dissipative coefficient term can be modified by the linear contribution from $\dot{\theta}(t)$ which is usually studied in the standard dissipative BJJ \cite{ch6_41b,ch6_41f,ch6_41g,ch6_46a}. However, the damping coefficient of the form $\dot{z}\sim\gamma Nz$ can be useful for the study of two-state system, such as, two condensates in different hyperfine levels in a single harmonic trap, connected by the tunneling transition \cite{ch6_46a}. In absence of dissipation ($\gamma=0$) the equations (\ref{eqn:imbalance}) and (\ref{eqn:phase}) reduce to the standard BJJ equations.\\

\subsection{Quantum and thermal properties of noise}

The noise properties of the operator $\hat F_{L,R}$ and $\hat F^{\dagger}_{L,R}$ in equations (\ref{eqn:20}) and (\ref{eqn:21}) can be derived using a suitable canonical thermal distribution of the bath operators at $t_{0}=0$. To this end we define the quantum statistical average of any reservoir operator $\hat O$ 
\begin{eqnarray}
\langle \hat O \rangle_{\rm qs}=\frac{{\rm Tr}[\hat O \exp{(-\hat H_R/k_{B}T)}]}{{\rm Tr}[\exp{(-\hat H_R/k_{B}T)}]}
\end{eqnarray}
where $\hat H_{R}=\sum_{j}\hbar\omega_{j}\hat n_{j}$ at $t_{0}=0$ and $\hat n_{j}$ denotes the number operator in $j$-th bath mode. Based on the above considerations the noise properties of the operator may be calculated using the canonical distribution of Eqs. (\ref{eqn:20}) and (\ref{eqn:21}). This immediately gives $\langle \hat F_{L,R}(t)\rangle_{\rm qs}=0$ and $\langle \hat F^{\dagger}_{L,R}(t)\rangle_{\rm qs}=0$ and $\langle \hat F^{\dagger}_{L,R}(t)\hat F_{L,R}(t')\rangle_{\rm qs}=\gamma\bar n_{L,R}(\Omega)\delta(t-t')$ and $\langle \hat F_{L,R}(t)\hat F^{\dagger}_{L,R}(t')\rangle_{\rm qs}=\gamma[\bar n_{L,R}(\Omega)+1]\delta(t-t')$. The fluctuation-dissipation relation gives 
\begin{widetext}
\begin{eqnarray}
\langle \hat F^{\dagger}_{L,R}(t)\hat F_{L,R}(t')+\hat F_{L,R}(t)\hat F^{\dagger}_{L,R}(t')\rangle_{\rm qs}&=&\gamma[2\bar n_{L,R}(\Omega)+1]\delta(t-t')\label{eqn:33}\\&=&\gamma \coth(\hbar\Omega/2k_{B}T)\delta(t-t')\nonumber
\end{eqnarray}
where the contangent hyperbolic factor in Eq. (\ref{eqn:33}) can be identified with Bose-Einstein distribution 
\begin{eqnarray}
\bar n_{L,R}(\Omega)=\frac{1}{e^{\hbar\Omega/k_{B}T} -1}
\end{eqnarray}
using the following relation $2\bar n_{L,R}(\Omega)+1=\coth(\hbar\Omega/2k_{B}T)$ and the plus one factor is responsible for the vacuum fluctuation which is always present on quantum scale even at absolute zero temperature.

Now to realize $\xi_{L,R}(t)$ as an effective $c$-number noise, we now introduce the ansatz that $\mu_{Lk,Rk}(0)$ and $\mu^{*}_{Lk,Rk}(0)$ are distributed according to Wigner thermal canonical distribution of Gaussian form \cite{ch6_40} as follows:
\begin{eqnarray}
W_{Lk,Rk}[\mu_{Lk,Rk}(0),\mu^{*}_{Lk,Rk}(0)]=N_{BL,BR}\exp \left[-\frac{|\mu_{Lk,Rk}(0)|^2}{2 \coth (\frac{\hbar \Omega}{2k_{B}T})} \right]
\label{eqn:35}
\end{eqnarray}
Here $N_{BL,BR}$ is the normalization constant for $L$ and $R$ wells. $\coth (\frac{\hbar \Omega}{2k_{B}T})$ is the width of the distribution. For any arbitrary quantum mechanical mean value of the bath operator $\langle \hat B_{Lk,Rk}\rangle$ which is a function of $\mu_{Lk,Rk}(0),\mu^{*}_{Lk,Rk}(0)$, its statistical average can then be calculated as 
\begin{eqnarray}
\langle\langle \hat B_{Lk,Rk}\rangle\rangle_{s}=\int \langle \hat B_{Lk,Rk}\rangle W_{Lk,Rk}[\mu_{Lk,Rk}(0),\mu^{*}_{Lk,Rk}(0)]d\mu_{Lk,Rk}(0)d\mu^{*}_{Lk,Rk}(0)
\label{eqn:36}
\end{eqnarray}
\end{widetext}
Using the ansatz (\ref{eqn:35}) and the definition of statistical average of Eq. (\ref{eqn:36}), one can show that $c$-number noise satisfy the following relations 
\begin{eqnarray}
\langle  \xi_{L,R}(t)\rangle_{s}=0\label{eqn:37}\\\langle  \xi^{*}_{L,R}(t)\rangle_{s}=0\nonumber
\end{eqnarray}
and
\begin{eqnarray}
\langle  \xi^{*}_{L,R}(t) \xi_{L,R}(t')\rangle_{s}=\gamma\coth (\hbar\Omega/2k_{B}T)\delta(t-t')\label{eqn:38}\\
\langle  \xi_{L,R}(t)\hat \xi^{*}_{L,R}(t')\rangle_{s}=\gamma\coth (\hbar\Omega/2k_{B}T)\delta(t-t')\nonumber
\end{eqnarray}
Eqs. (\ref{eqn:37}) and (\ref{eqn:38}) implies that the $c$-number noise $\xi_{L,R}(t)$ is characterized by zero mean and follows the fluctuation-dissipation relation. By using the $c$-number formalism, we may thus bypass the operator ordering prescription for the derivation of noise properties. The $c$-number noise  $\xi_{L,R}(t)$ as characterized by Eqs. (\ref{eqn:37}) and (\ref{eqn:38}) is classical looking in form but essentially quantum mechanical in nature \cite{ch6_47}.

\section{Phase Diffusion}\label{sec:3}

The Eqs. (\ref{eqn:imbalance}) and (\ref{eqn:phase}) describe the dissipative BJJ equations with quantum noise whose properties are governed by Eqs. (\ref{eqn:37}) and (\ref{eqn:38}). These are nonlinear Langevin equations which can not be solved by any direct analytical method. The traditional way to circumvent this difficulty is to take resort to the weak noise limit. To this end we consider first the steady state of the system in absence of noise and linearize the dynamics around it. We are then led a multivariate Ornstein–Uhlenbeck (OU) process as considered below.

To proceed further, we begin with the steady state of the  dynamical system ($z_s,\theta_{s}$). Now linearizing the system around it  with $z=z_{s}+\delta z$ and $\theta=\theta_{s}+\delta\theta$, where $\delta z$ and $\delta \theta$ are small perturbations, we obtain the linearized Langevin equations in $c$-numbers
\begin{widetext}
\begin{equation}
 \delta\dot{z}=-\frac{2K}{\hbar}{\sqrt{1-z_{s}^2}}\cos\theta_{s}\delta\theta+\frac{2K}{\hbar}\frac{z_{s}\delta z}{\sqrt{1-z_{s}^2}}\sin\theta_{s} -\gamma N\delta z +\xi_{z} (t)
 \label{eqn:39}
\end{equation}
\begin{equation}
 \delta\dot{\theta}=\frac{2K}{\hbar}\left[\Lambda_0\delta z-\frac{z_{s}\sin\theta_{s}}{\sqrt{1-z_{s}^2}}\delta\theta+\frac{\cos\theta_{s}}{(1-z_{s}^2)^{\frac{3}{2}}}\delta z \right]+\xi_{\theta} (t)
 \label{eqn:40}
\end{equation}
where $\xi_{z} (t)= \xi_{L}(t)-\xi_{R}(t)+z_{s}[\xi_{L}(t)+\xi_{R}(t)]$ and $\xi_{\theta} (t)=\xi_{R}(t)-\xi_{L}(t)$. Here, we have used the relations $N_{Ls,Rs}=\frac{N}{2}(1 \pm z_{s})$. $N_{Ls,Rs}$ are the steady state values of the number of atoms in left and right wells. For $\gamma>0$, $\delta z$ gets equilibrated at a fast rate. Therefore adiabatic elimination of the fast variable $z$ results in :  
\begin{eqnarray}
\delta z=\frac{\frac{2K}{\hbar}\sqrt{1-z_{s}^2}\cos\theta_{s}\delta \theta-\xi_{z}}{\frac{2K}{\hbar} \frac{z_{s}\sin\theta_{s}}{\sqrt{1-z_{s}^2}}-\gamma N}
\end{eqnarray}
Now inserting $\delta z$ in Eq. (\ref{eqn:40}), we finally obtain after some algebra an equation in the form of linear Langevin dynamics for phase $\delta \theta (t)$
\begin{eqnarray}
\delta\dot{\theta}=-\eta \delta \theta +C\xi_{z}(t)+\xi_{\theta}(t)
\label{eqn:42}
\end{eqnarray}
where $\eta$ is the linear phase drift. The general expressions for $\eta$ and $C$ are 
\begin{eqnarray}
\eta=\frac{2K}{\hbar}\left[\frac{z_{s}\sin\theta_{s}}{\sqrt{1-z_{s}^2}}-\frac{\cos\theta_{s}}{(1-z_{s}^2)^{\frac{3}{2}}}\left\{\frac{\sqrt{1-z_{s}^2}\cos\theta_{s}}{\frac{z_{s}\sin\theta_{s}}{\sqrt{1-z_{s}^2}}-\frac{\gamma N\hbar}{2K}}\right\}-\Lambda_0\left\{\frac{\sqrt{1-z_{s}^2}\cos\theta_{s}}{\frac{z_{s}\sin\theta_{s}}{\sqrt{1-z_{s}^2}}-\frac{\gamma N\hbar}{2K}}\right\}\right]
\end{eqnarray}
and
\begin{eqnarray}
C=-\frac{\cos\theta_{s}}{(1-z_{s}^2)^{\frac{3}{2}}}\left\{\frac{1}{\frac{z_{s}\sin\theta_{s}}{\sqrt{1-z_{s}^2}}-\frac{\gamma N\hbar}{2K}}\right\}-\Lambda_0\left\{\frac{1}{\frac{z_{s}\sin\theta_{s}}{\sqrt{1-z_{s}^2}}-\frac{\gamma N\hbar}{2K}}\right\}
\label{eqn:C}
\end{eqnarray}
\end{widetext}
,respectively. In the subsequent sections, we will use the expressions of $\eta$ and $C$ to calculate the phase diffusion coefficient for zero-phase mode ($z_{s}=0, \theta_{s}=0$) and $\pi$-phase modes ($z_{s}=0, \theta_{s}=\pi$; $z_{s}=\pm \sqrt{1-\frac{1}{\Lambda_{0}^2}}, \theta_{s}=\pi$).
Making use of Eqs. (\ref{eqn:37}) and (\ref{eqn:38}) one can show 
\begin{eqnarray}
\hspace{-0.5cm}
\langle \xi(t) \xi(t')\rangle=C^2(1+z_{s}^2) 2\gamma \coth (\hbar \Omega/2k_{B}T)\delta(t-t')
\end{eqnarray}
where  $\xi(t)=C \xi_{z}(t)+ \xi_{\theta}(t)$.
The Fokker-Planck equation corresponding to the linear Langevin dynamics (Eq. (\ref{eqn:42})) is given by \cite{ch6_43}
\begin{eqnarray}
\frac{\partial P(\Psi,t)}{\partial t}=-\frac{\partial}{\partial \Psi} (\eta \Psi) P(\Psi,t)+\mathcal{D}\frac{\partial^2P(\Psi,t)}{\partial \Psi^2}  
\label{eqn:46}
\end{eqnarray}
Here $P(\Psi,t)$ is the probability of finding $\Psi$ at time $t$; we have put $\delta \theta \equiv \Psi$. $\mathcal{D}$ is the phase diffusion coefficient as given by
\begin{eqnarray}
\mathcal{D}= C^2\gamma(1+z_{s}^2) \coth (\hbar \Omega/2k_{B}T)
\label{eqn:47}
\end{eqnarray}
A clear separation of the statistical part $\coth (\hbar \Omega/2k_{B}T)$ from the dynamical pre-factor is quite apparent in the above expression.
The phase diffusion coefficient of Eq. (\ref{eqn:47}) is one of the main results of this section. Eq. (\ref{eqn:46}) shows that phase perturbations can be described as Brownian motion of a particle that are characterized by phase drift and diffusion. At finite temperature, the macroscopic quantum tunneling across 1D dissipative BJJ is thus significantly affected by number- and phase- fluctuations.

Before closing this section, we digress a little bit about the consistency check of the calculation. Putting the Fokker-Planck Eq. (\ref{eqn:46}) in the form of a continuity equation $\frac{\partial P}{\partial t}+\frac{\partial F}{\partial \Psi}=0$, we identify the flux \cite{ch6_43}
\begin{eqnarray}
F=-\left[\eta \Psi P+\mathcal{D}\frac{\partial P}{\partial \Psi}\right]
\label{eqn:48}
\end{eqnarray}
At equilibrium $F=0$, we obtain from Eq. (\ref{eqn:48}) the equilibrium distribution function in the zero-phase mode, $\pi$-phase mode ($z_{s}=0, \theta_{s}=\pi$) and $\pi$-phase self-trapping mode ($z_{s}=\pm \sqrt{1-\frac{1}{\Lambda_{0}^2}}, \theta_{s}=\pi$)
\begin{eqnarray}
P_{0}(\Psi)= A \exp \left[-\left\{\frac{N(1+\Lambda_0)^{-1}}{2\coth (\frac{\hbar \Omega}{2k_{B}T})}\right\} \Psi^2\right]
\label{eqn:49}
\end{eqnarray}

\begin{eqnarray}
P_{\pi}(\Psi)= A \exp \left[-\left\{\frac{N(1-\Lambda_0)^{-1}}{2\coth (\frac{\hbar \Omega}{2k_{B}T})}\right\} \Psi^2\right]
\label{eqn:49a}
\end{eqnarray}

\begin{eqnarray}
P_{\rm ST}(\Psi)= A \exp \left[-\left\{\frac{N[\Lambda_0^2(\Lambda_0^2-1)]^{-1}}{2\coth (\frac{\hbar \Omega}{2k_{B}T})}\right\} \Psi^2\right]
\label{eqn:49b}
\end{eqnarray}

at a finite temperature $T$ where $A$ is the normalization constant, respectively. The distribution does not depend on $\gamma$ as it should be for the attainment of equilibrium. Second, the width is governed by $\coth$-function which carries the signature of Wigner canonical thermal distribution (\ref{eqn:35}) employed in course of construction of ensemble for $c$-number noise in the present treatment. Third, the occurrence of the interaction parameter $\Lambda_{0}$ in the $\Psi^2$ term of the parenthesis of the distributions of Eqs. (\ref{eqn:49}), (\ref{eqn:49a}) and (\ref{eqn:49b})  for both zero and $\pi$-phase modes is reminiscent of the potential energy term of an equilibrium distribution. 

\section{Spectrum of fluctuations}\label{sec:4}

We now go beyond the adiabatic elimination of fast variable to calculate the spectra of fluctuation of number imbalance and phase associated with the dissipative BJJ dynamics. To proceed we recast Eqs. (\ref{eqn:39}) and (\ref{eqn:40}) in the matrix form as follows
\begin{eqnarray}
\dot \beta(t) = -B \beta(t) + M(t)
\label{eqn:50}
\end{eqnarray}
where
\begin{widetext}
\begin{equation}
\beta=
\begin{pmatrix}
\delta z \\
\delta \theta  
\end{pmatrix}, \\
B=
\begin{pmatrix}
\gamma N -\frac{2K}{\hbar}\frac{z_{s}\sin\theta_{s}}{\sqrt{1-z_{s}^2}} &  \frac{2K}{\hbar}{\sqrt{1-z_{s}^2}}\cos\theta_{s}  \\

-\frac{2K}{\hbar}\left\{\Lambda_0+\frac{\cos\theta_{s}}{(1-z_{s}^2)^{\frac{3}{2}}}\right\}  & \frac{2K}{\hbar}\frac{z_{s}\sin\theta_{s}}{\sqrt{1-z_{s}^2}}
\end{pmatrix}, \\
M=
\begin{pmatrix}
\xi_{z} \\
\xi_{\theta} 
\end{pmatrix}
\end{equation}
\end{widetext}
such that $\beta \beta^{T}\equiv \beta \times \beta$ is a direct product of the matrix
\begin{equation}
\beta \times \beta=
\begin{pmatrix}
\delta z(t) \delta z(t) &  \delta z(t) \delta \theta(t)  \\
\delta \theta(t) \delta z(t) & \delta \theta(t) \delta \theta(t) 
\end{pmatrix}
\end{equation}
Taking average on the both sides of the Eq. (\ref{eqn:50}) we obtain
\begin{equation}
\langle \dot \beta (t) \rangle= -B \langle \beta(t) \rangle +\langle M(t) \rangle
\end{equation}
By virtue of Eq. (\ref{eqn:37}), we have $\langle M(t) \rangle=0$. Direct integration yields
\begin{eqnarray}
\langle \beta (t) \rangle= e^{-B t} \langle \beta(0) \rangle
\end{eqnarray}
where $\langle \beta(0) \rangle$ gives the average of the initial value $\langle \beta(t) \rangle$. Now, according to the regression theorem the correlation function decays in the same way as the average decay which suggests that
\begin{eqnarray}
\langle \beta (t)\beta (0) \rangle= e^{-B t} \langle \beta(0)\beta(0) \rangle \label{eqn:55}\\ 
\langle \beta (0)\beta (t) \rangle= \langle \beta(0)\beta (0) \rangle e^{-B^{T} t} \nonumber
\end{eqnarray}
To obtain the low frequency spectrum of various modes of correlation, we calculate the Fourier transform of $\langle \beta (t)\beta (0) \rangle$ and define,
\begin{eqnarray}
\mathcal{S}(\Delta)=\int_{-\infty}^{+\infty}e^{-i\Delta I t} \langle \beta (t)\beta(0) \rangle dt
\label{eqn:56}
\end{eqnarray}
where, $\Delta$ refers to the detuning around $\Omega$, $I$ is the identity matrix and $t=0$ refers to the stationary state, i.e, we calculate the correlation of fluctuation around the stationary state. On further manipulation of Eqs. (\ref{eqn:56}) and (\ref{eqn:55}), we obtain
\begin{widetext}
\begin{eqnarray}
\mathcal{S}(\Delta)=(B+i\Delta I)^{-1}\langle \beta (0)\beta(0) \rangle + \langle \beta (0)\beta(0) \rangle (B^{T}-i\Delta I)^{-1}
\end{eqnarray}
The stationary state contribution in the above equation can be expressed in terms of the diffusion matrix of the form
\begin{eqnarray}
\mathcal{D}_{z,\theta}=
\begin{pmatrix}
\gamma (1+z_{s}^2)\coth(\hbar\Omega/2k_{B}T) & 0 \\
0 & \gamma \coth(\hbar\Omega/2k_{B}T) 
\end{pmatrix}
\end{eqnarray}
so that $2\times 2$ fluctuation spectrum matrix becomes
\begin{eqnarray}
\mathcal{S}(\Delta)=(B+i\Delta I)^{-1}2\mathcal{D}_{z,\theta} (B^{T}-i\Delta I)^{-1}
\end{eqnarray}
Explicit evaluation of the matrix elements results in the fluctuations of number imbalance as $\mathcal{S}_{11}$ element, while $\mathcal{S}_{22}$ element represents the contribution due to phase fluctuation \cite{ch6_42}.

The number fluctuation spectrum ($\mathcal{S}_{11}$ element)  can be written as
\begin{eqnarray}
\mathcal{S}_{z}(\Delta)=\frac{2\mathcal{D}_{11}(B_{22}^2+\Delta^2)+2\mathcal{D}_{22}B_{12}^2}{(B_{11}B_{22}-B_{12}B_{21}-\Delta^2)^2+\Delta^2(B_{11}+B_{22})^2}
\label{eqn:58}
\end{eqnarray}

Similarly, the phase fluctuation term ($\mathcal{S}_{22}$ element) can be written as
\begin{eqnarray}
\mathcal{S}_{\theta}(\Delta)=\frac{2\mathcal{D}_{11}B_{21}^2+2\mathcal{D}_{22}(B_{11}^2+\Delta^2)}{(B_{11}B_{22}-B_{12}B_{21}-\Delta^2)^2+\Delta^2(B_{11}+B_{22})^2}
\label{eqn:59}
\end{eqnarray}
\end{widetext}
where, $B_{11}=\gamma N -\frac{2K}{\hbar}\frac{z_{s}\sin\theta_{s}}{\sqrt{1-z_{s}^2}}$, $B_{12}=\frac{2K}{\hbar}{\sqrt{1-z_{s}^2}}\cos\theta_{s}$, $B_{21}=-\frac{2K}{\hbar}\left\{\Lambda_0+\frac{\cos\theta_{s}}{(1-z_{s}^2)^{\frac{3}{2}}}\right\}$, and $B_{22}=\frac{2K}{\hbar}\frac{z_{s}\sin\theta_{s}}{\sqrt{1-z_{s}^2}}$. The diffusion matrix elements are defined as
$\mathcal{D}_{11}=\gamma (1+z_{s}^2)\coth(\hbar\Omega/2k_{B}T)$ and
$\mathcal{D}_{22}=\gamma \coth(\hbar\Omega/2k_{B}T)$.

\section{Results and discussions}\label{sec:5}
\subsection{Coherence factor}
In order to study the coherent and incoherent regimes of the dissipative BJJ, we now define $\langle \cos\Psi\rangle$ as a coherence factor \cite{ch6_41} as it provides the degree of coherence of the system. If the value of the linearized phase is localized around zero, the value of the coherence factor is close to unity. If instead the phase is fully delocalized and all its values are equally probable, then the value of the coherence factor is close to zero implying that the system is in the incoherent state. Since at equilibrium, the linearized phase follow the Wigner thermal canonical distribution \cite{ch6_40} as described by Eqs. (\ref{eqn:49}), (\ref{eqn:49a}) and (\ref{eqn:49b}), one may define explicitly the coherence factor as follows: 
\begin{eqnarray}
\langle \cos\Psi \rangle=\frac{\int_{-\pi}^{\pi}d\Psi \cos \Psi \exp\left[{\frac{\cos\Psi}{\kappa\coth(\frac{\hbar \Omega}{2k_{B}T})} }\right] }{\int_{-\pi}^{\pi}d\Psi\exp\left[{\frac{\cos\Psi}{\kappa\coth(\frac{\hbar \Omega}{2k_{B}T})} }\right]}
\label{eqn:50new}
\end{eqnarray}
where $\kappa$ is defined as $\kappa=\frac{1+\Lambda_{0}}{N}$ for zero-phase mode. The appearance of the factor $\kappa$ in the canonical distribution makes the coherence factor dependent on the on-site interaction energy $U$ and the tunneling energy $K$. In what follows we examine the coherence factor in the light of these parameters. For our numerical calculation, we choose the value of the total number of atoms $N=1000$.

\begin{figure}[h]
\centering
\begin{tabular}{c}
\includegraphics[width=4.0cm]{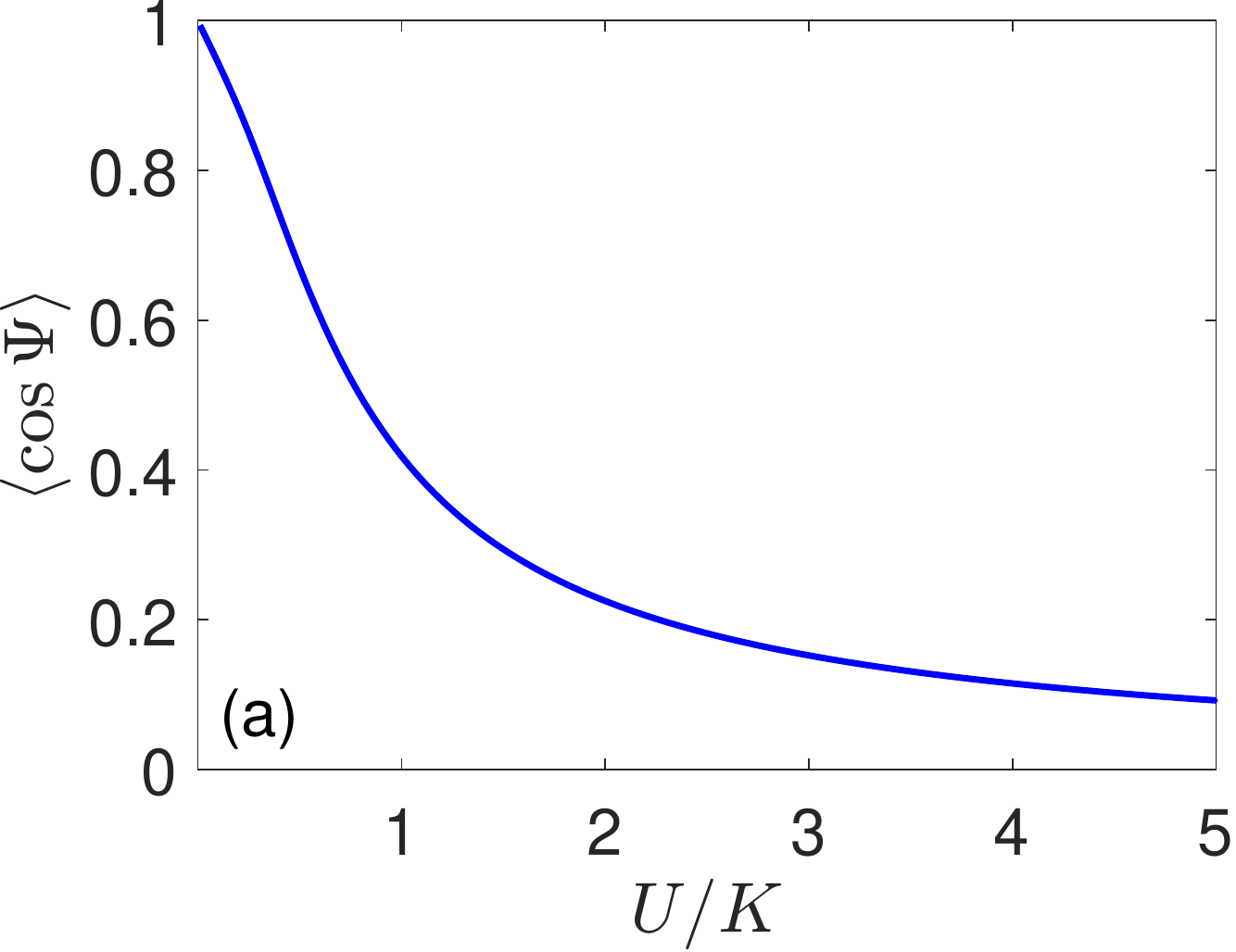}
\includegraphics[width=4.0cm]{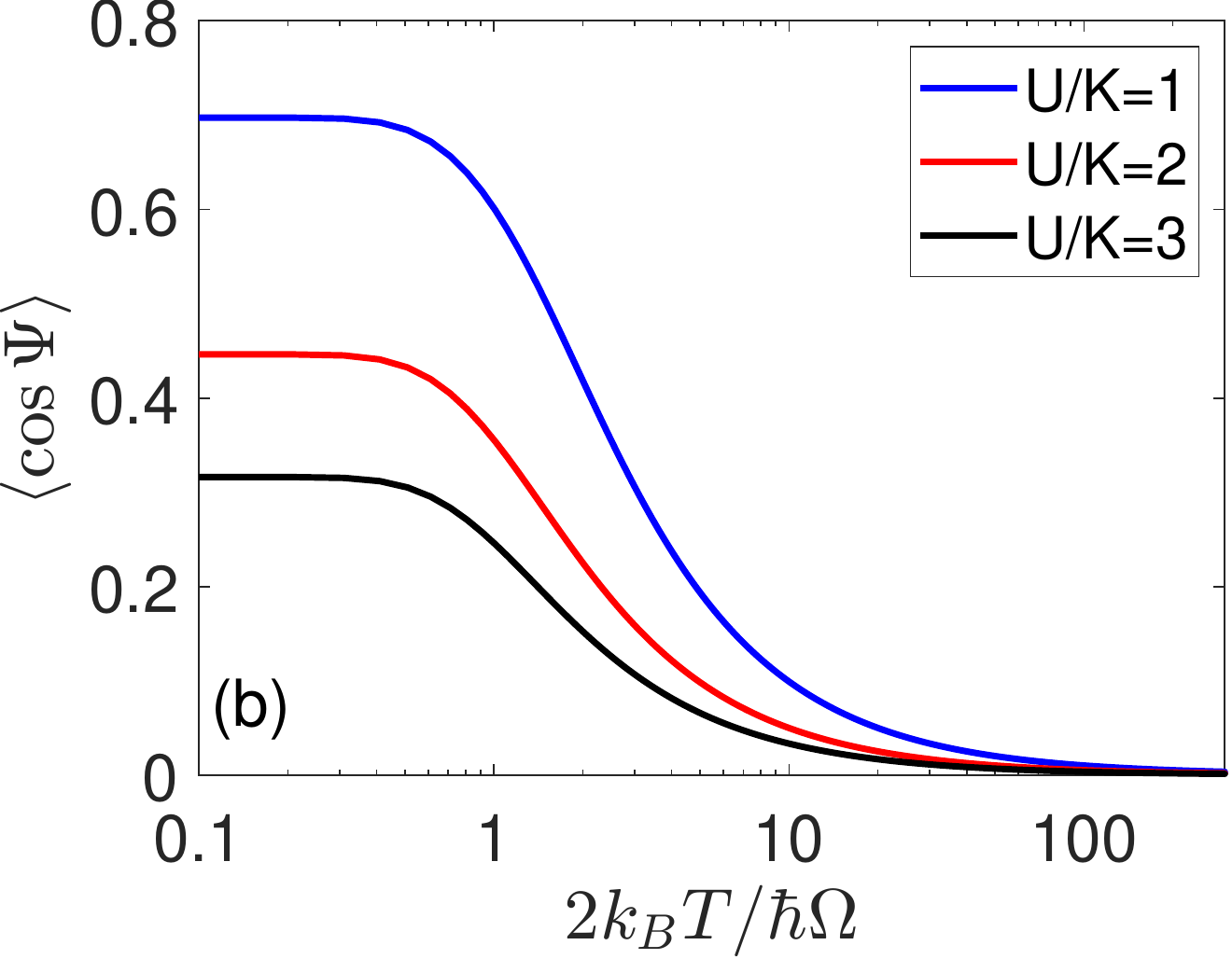}
\end{tabular}
\caption{(Color online) (a) Variation of coherence factor $\langle \cos \Psi \rangle$ as a function of $U/K$ at low temperature $k_{B}T/\hbar\Omega=1/10$ in zero-phase mode. (b) Variation of coherence factor $\langle \cos \Psi \rangle$ as a function of $2k_{B}T/\hbar\Omega$ for different energies $U/K=1$ (blue solid line), $U/K=2$ (red solid line), $U/K=3$ (black solid line). Note the logarithmic scale on the horizontal axis.}
\label{fig:1a}
\end{figure}

In Fig. \ref{fig:1a}(a), we show that the coherence factor $\langle \cos \Psi \rangle$ as a function of the ratio $U/K$ for the low temperature limit $k_{B}T/\hbar\Omega=1/10$ in zero-phase mode. It is apparent that in the limit of strong tunneling $U/K\ll 1$, the value of the coherence factor becomes close to unity because under this condition the system undergoes small oscillation around the equilibrium zero-phase value. In this limit the fluctuations of the phase is also small. In the opposite limit, when $U/K\gg1$, the amplitude of the oscillation around the equilibrium increases as a result of delocalization of the linearized phase due to the large on-site interaction. The phase fluctuations are not small and the coherence factor gradually decreases. This is similar to the prediction of the coherence factor calculated using Josephson Hamiltonian \cite{ch6_41}.

In Fig. \ref{fig:1a}(b), we plot the temperature dependence of the coherence factor for fixed interaction energies in zero-phase mode. It is clear that for a particular value of the interaction energy, the coherence factor for low temperature is almost constant and then with increase of the temperature the coherence factor decreases. At high temperature ($2k_{B}T/\hbar\Omega\approx 100$), the curves coincide for different interaction energies and becomes close to zero which implies that at large temperature the system becomes incoherent. It is also noted that the slope of the coherence factor decreases with increase of the on-site interaction energy which implies that in the limit of negligible tunneling $U/K\gg1$ for a particular temperature the coherence factor decreases which is consistent with Fig. \ref{fig:1a}(a). This general behaviour of the coherence factor is observed over a three orders of magnitude of $2k_{B}T/\hbar\Omega$, and is in good agreement with the experimental observation \cite{ch6_39}.

\begin{figure}[h]
\centering
\begin{tabular}{c}
\includegraphics[width=4.0cm]{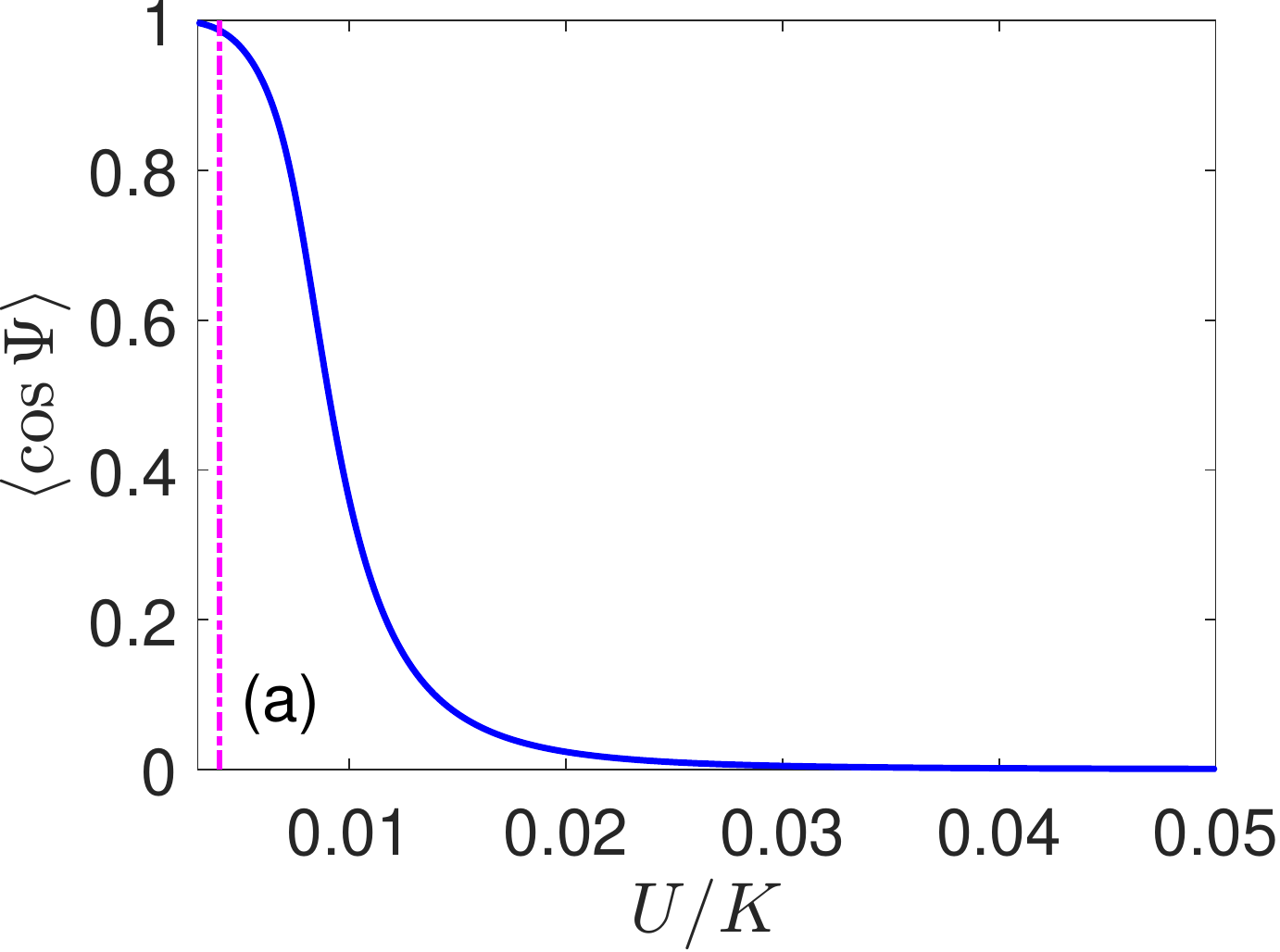}
\includegraphics[width=4.0cm]{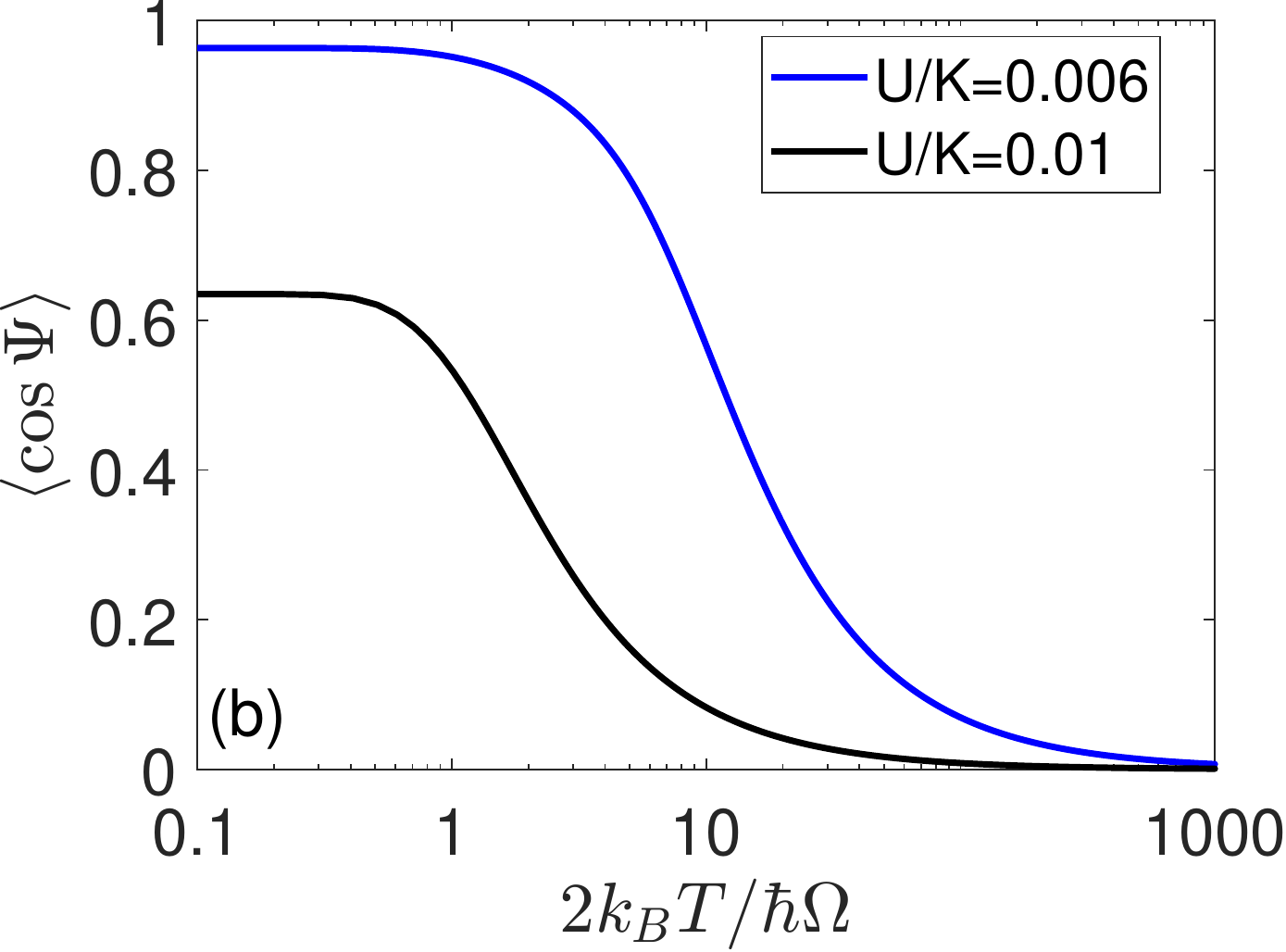}
\end{tabular}
\caption{(Color online) (a) Variation of coherence factor $\langle \cos \Psi \rangle$ as a function of $U/K$ at low temperature $k_{B}T/\hbar\Omega=1/10$ in the $\pi$-phase self-trapping regime. The vertical dashed-dotted magenta line corresponds to $U/K=0.004$. See text for details. (b) Variation of coherence factor $\langle \cos \Psi \rangle$ as a function of $2k_{B}T/\hbar\Omega$ for different energies $U/K=0.006$ (blue solid line), $U/K=0.01$ (black solid line). Note the logarithmic scale on the horizontal axis.}
\label{fig:1anew}
\end{figure}

In the $\pi$-phase mode there are two possible regimes where one can study the effect of coherence with the help of the many-body interaction parameter $\Lambda_{0}$. The regime $\Lambda_{0}<1$ corresponds to the $\pi$-phase Josephson oscillation regime and $\Lambda_{0}>1$ refers to the $\pi$-phase self-trapping regime. The regime $1<\Lambda_{0}<2$ specifies the $\pi$-phase self-trapping mode and $\Lambda_{0}>2$ refers to the running phase self-trapping regime which is similar to the zero-phase mode self-trapping regime. We have checked that in the limit of $\pi$-phase Josephson oscillation regime the value of $\kappa$ is defined as $\kappa=\frac{1-\Lambda_{0}}{N}$ for $\Lambda_{0}<1$ which implies that $U/K\ll1$ and as a result the value of the coherence factor is positive and lies in between zero and unity. However, the nature of the coherence factor as a function of $U/K$ and temperature remains same as in zero-phase mode.

In Fig. \ref{fig:1anew}(a), we plot the coherence factor $\langle \cos \Psi \rangle$ as a function of the ratio $U/K$ for the low temperature limit $k_{B}T/\hbar\Omega= 1/10$ in $\pi$-phase self-trapping regime. In this regime $\kappa$ is defined as $\kappa=\frac{\Lambda_{0}^2[\Lambda_{0}^2-1]}{N}$ with $\Lambda_{0}>1$. We see that the coherence factor quickly falls from unity to zero with small change in the $U/K$ ratio. We also see that when $U/K<0.004$, the value of the coherence factor is close to unity. This regime signifies the $\pi$-phase self-trapping regime where the system oscillates around a non zero-value of the population imbalance and phase difference. Also in this regime the system undergoes oscillation around the equilibrium $\pi$-phase value and the fluctuation of the phase remains low. We also note that when $U/K>0.004$ the coherence factor suddenly falls from unity to zero. In Fig. \ref{fig:1anew}(a), the vertical dashed-dotted magenta line corresponds to $U/K=0.004$ which basically separates the $\pi$-phase self-trapping regime to running $\pi$-phase self-trapping regime. For $U/K>0.004$, in the running $\pi$-phase self-trapping regime where the population imbalance oscillates around a non-zero mean value but the phase difference between the two BECs in the left and right well evolves unbound as a result of which the relative phase increases monotonically. So, the phase is delocalized and the phase fluctuations are not small. The system becomes incoherent. We also see that when $U/K>0.025$, the value of the coherence factor is almost zero which signifies that the system is in the incoherent regime.

In Fig. \ref{fig:1anew}(b), we show that the temperature dependence of coherence factor for fixed interaction energies in $\pi$-phase self-trapping regime. To study this aspect we choose the value of $U/K$ in the nearly coherent regime where the system is in the running phase self-trapping mode. We observe similar behaviour when compared to the case of zero phase mode where in the low temperature limit the coherence factor is almost constant. Then with increase of the temperature the coherence factor decreases and also at high temperature ($2k_{B}T/\hbar\Omega\approx 1000$), the curves coincide for the different interaction energies and become close to zero which implies that the system is in the incoherent regime. However the degree of coherence increases by one order of magnitude compared to the zero-phase mode. This study of coherence reveals that the 1D dissipative BJJ has a higher degree of coherence.

\subsection{Phase diffusion coefficient; phase-transition-like behaviour}
\begin{figure}[h]
\centering
\begin{tabular}{c}
\includegraphics[width=7.1cm]{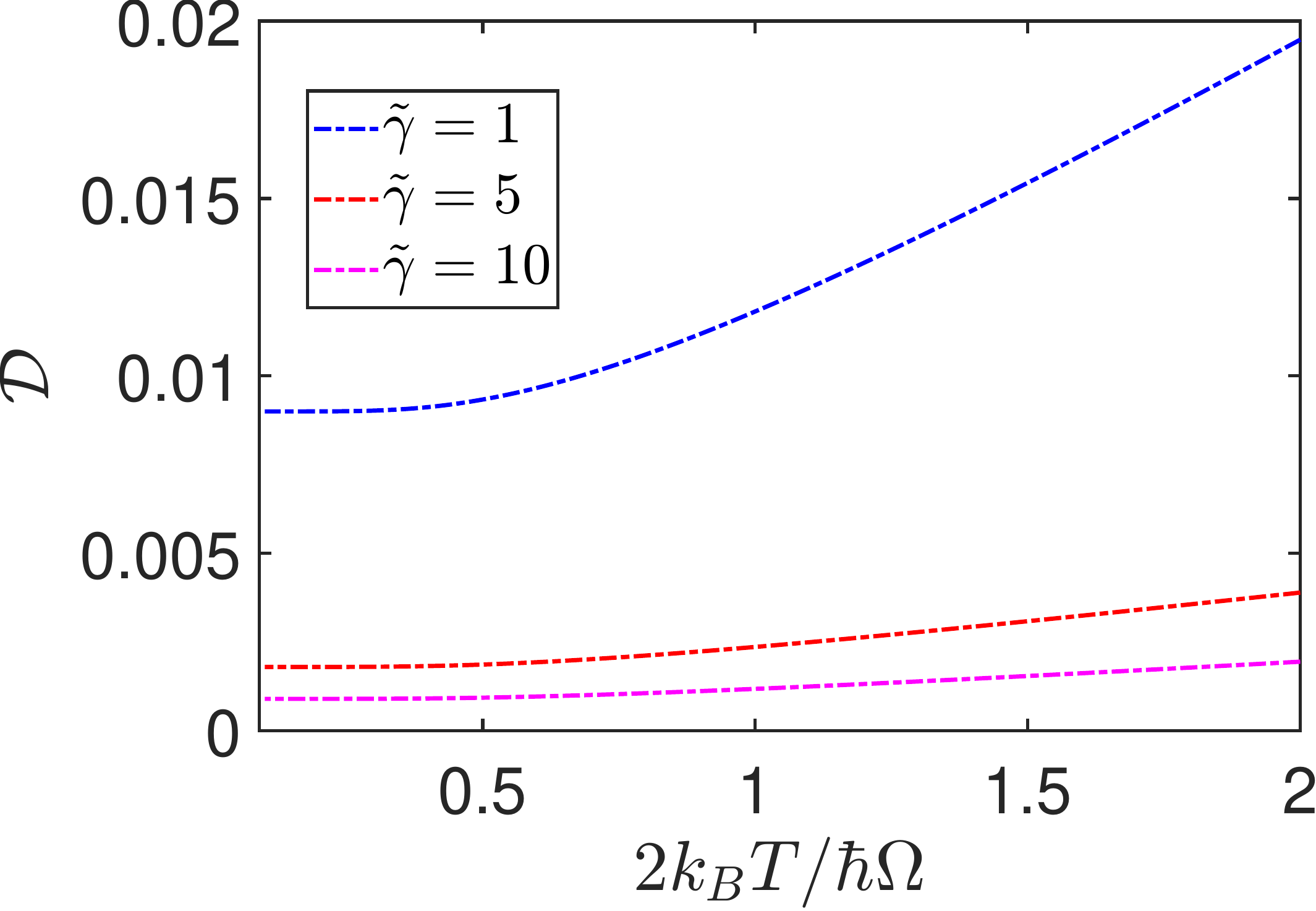}
\end{tabular}
\caption{(Color online) Variation of the phase diffusion co-efficient $\mathcal{D}$ as a function of temperature $2k_{B}T/\hbar\Omega$ for different dissipation co-efficient $\tilde{\gamma}$ with small many body interaction parameter $\Lambda_{0}=0.5$ in zero-phase mode.}
\label{fig:1}
\end{figure}

Here we present our results for the phase diffusion coefficient in zero-phase mode and $\pi$-phase modes of 1D dissipative BJJ. We first analyze how phase diffusion coefficient $\mathcal{D}$ depends on the system interaction parameter and temperature. For this we choose the zero-phase mode, i.e $z_s=0,\theta_s=0$ which is well studied in BJJ. From equation (\ref{eqn:47}), in zero-phase mode, the phase diffusion coefficient is given by
\begin{eqnarray}
\mathcal{D}= C^2\gamma \coth (\hbar \Omega/2k_{B}T)
\end{eqnarray}
where $C$ in the zero-phase mode is given by $C=\frac{2K}{\hbar\tilde{\gamma}}\left[1+\Lambda_{0}\right]$ where $\tilde{\gamma}=\gamma N$. A closer look at the analytical expression makes it clear that $\mathcal{D}\propto1/\gamma$ (as $C^2\propto1/\tilde{\gamma}^2$). To investigate how the phase diffusion coefficient $\mathcal{D}$ changes with temperature for weak to strong dissipation limit, we plot $\mathcal{D}$ as a function of $2k_{B}T/\hbar\Omega$ for several dissipation coefficient $\tilde{\gamma}$ in Fig. \ref{fig:1}. For numerical calculation, we choose the total number of atoms $N=1000$ and $U/K=0.001$ such that the many body interaction parameter becomes $\Lambda_{0}=0.5$, i.e, weak interaction limit or the strong tunneling regime. The reason behind this choice of parameter is that we want to investigate the nature of the phase diffusion coefficient in the coherent regime ($U/K\ll1$) by introducing dissipation in the system. Fig. \ref{fig:1} reveals that the $\mathcal{D}$ remains constant upto $2k_{B}T/\hbar\Omega \approx 0.4$ which is determined by the system parameters $\Lambda_0$ and $\tilde{\gamma}$. However, if we further increase the temperature i.e ($2k_{B}T/\hbar\Omega> 0.4$) $\mathcal{D}$ increases but for lower value of $\tilde{\gamma}$ as shown by the blue dashed-dotted line in Fig. \ref{fig:1}. If $\tilde{\gamma}$ is comparatively large the change of $\mathcal{D}$ with temperature occurs slowly which are shown by the magenta and red dashed-dotted lines in Fig. \ref{fig:1}. Since $\mathcal{D}$ is proportional to $C^2$ which is determined by the many body interaction parameter $\Lambda_{0}$, the increase in $\Lambda_0$ does not change effectively the nature of the phase diffusion coefficient.

\begin{figure}[h]
\centering
\begin{tabular}{c}
\includegraphics[width=7cm]{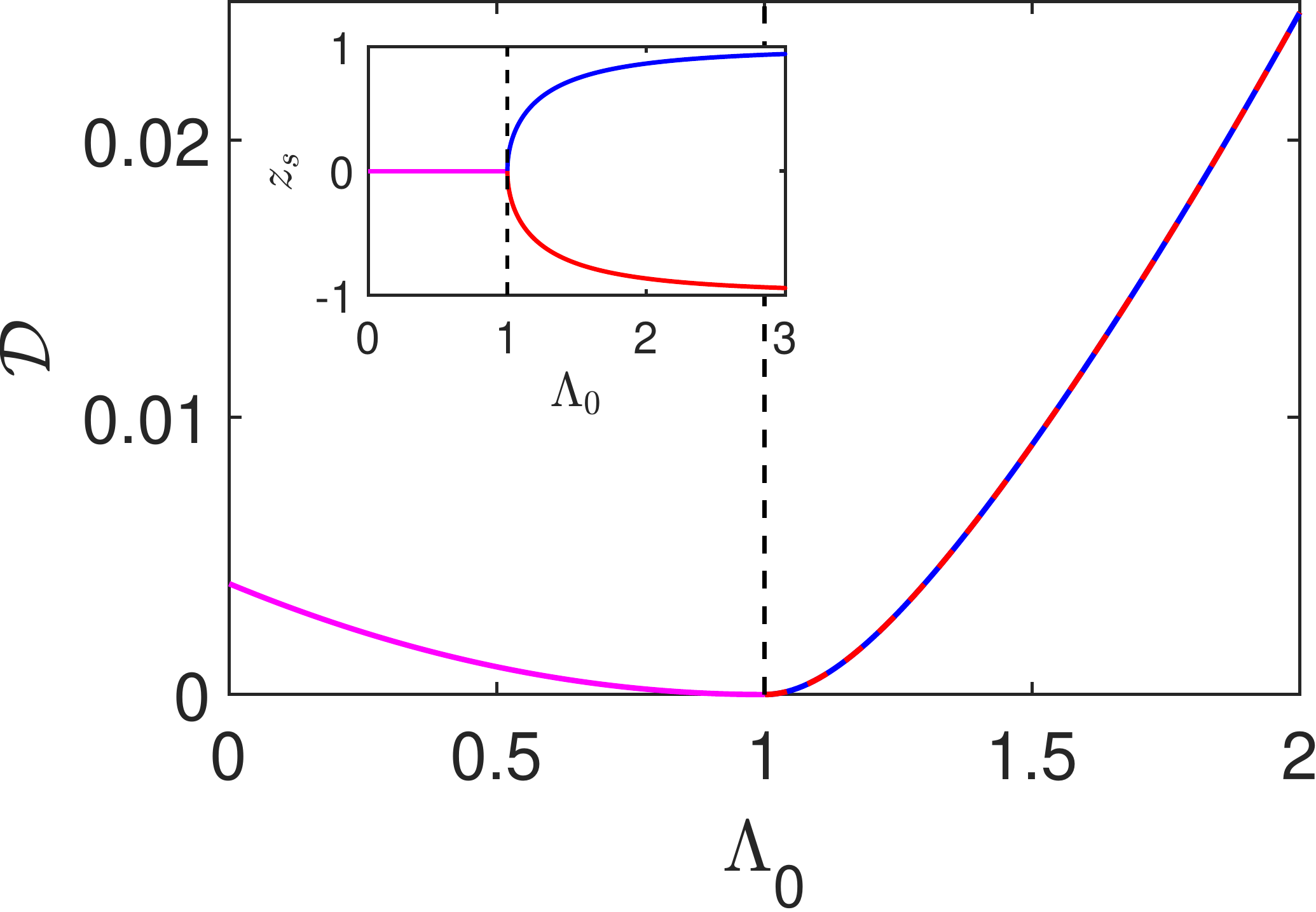}
\end{tabular}
\caption{(Color online) Variation of the phase diffusion coefficient $\mathcal{D}$ as a function of the many body interaction parameter $\Lambda_{0}$ at low temperature $k_{B}T/\hbar\Omega=1/10$ with $\tilde{\gamma}=1$ in the $\pi$-phase mode. The phase diffusion coefficient $\mathcal{D}$ for $z_{s}=0,\theta_{s}=\pi$ represented by magenta solid line and for $z_{s}=\pm\sqrt{1-\frac{1}{\Lambda_{0}^2}}, \theta_{s}=\pi$, $\mathcal{D}$ as a function of $\Lambda_{0}$ represented by solid blue and red dashed lines. The vertical black dashed line represent the critical $\Lambda_{0}$ for the transition between Josephson oscillation to MQST in $\pi$-phase mode.  
Inset: Variation of the steady state values of the population imbalance as a function of the many body interaction parameter $\Lambda_{0}$. The three different colors signify three different steady states of the population imbalance. The vertical black dashed line represents $\Lambda_{0}$ where the bifurcation of $z_{s}$ occurs.}
\label{fig:1_new}
\end{figure}

In the $\pi$-phase mode, there are three steady state solutions (a) $z_{s}=0$, $\theta_{s}=\pi$
(b) $z_{s}=\sqrt{1-\frac{1}{\Lambda_{0}^2}}$, $\theta_{s}=\pi$ and (c) $z_{s}=-\sqrt{1-\frac{1}{\Lambda_{0}^2}}$, $\theta_{s}=\pi$. The first solution arises for $\Lambda_{0}<1$ and the second and third solutions arise when $\Lambda_{0}>1$ and at $\Lambda_{0}=1$ a bifurcation of population imbalance occurs which is plotted in the inset of Fig. \ref{fig:1_new}. This bifurcation in the $\pi$-phase mode has been observed experimentally in internal BJJ \cite{ch6_47a}. To analyze the steady states in terms of the phase diffusion co-efficient, we plot the phase diffusion coefficient as a function of the interaction parameter $\Lambda_{0}$ in Fig. \ref{fig:1_new}. We see that for $\Lambda_{0}<1$, i.e, when the steady state of the population imbalance is expressed as $z_{s}=0$, $\theta_{s}=\pi$, the phase diffusion coefficient $\mathcal{D} \propto C^2\gamma$ where $C=\frac{2K}{\hbar\tilde{\gamma}}\left[\Lambda_{0}-1\right]$. The phase diffusion coefficient behaves similarly with dissipation ($\mathcal{D}\propto1/\gamma$) as in zero-phase mode. But there is a critical value of $\Lambda_{0}$ for which $C$ and $\mathcal{D}$ become zero. This phase-transition like behaviour originates due to the symmetry breaking of BJJ in the $\pi$-phase mode.  So with increase of $\Lambda_{0}$, the value of $C$ decreases and as a result the phase diffusion coefficient $\mathcal{D}$ decreases as represented by the solid magenta line in Fig. \ref{fig:1_new}. At $\Lambda_{0}=1$, the value of $C$ and in turn the phase diffusion coefficient becomes zero. But for $\Lambda_{0}>1$, there are two possible solutions of $z_{s}$ and for both cases $C$ is given by $C=\frac{2K}{\hbar\tilde{\gamma}}\left[\Lambda_{0}-\frac{1}{\Lambda_{0}^3}\right]$. For this, we see that $\mathcal{D}$ increases sharply with increase of $\Lambda_{0}$ represented by the blue solid line and red dashed line in Fig. \ref{fig:1_new}. However, the bifurcation shown for the steady state population imbalance is missing for the phase diffusion coefficient in the regime $\Lambda_{0}>1$. This can be understood from the phase diffusion equation (\ref{eqn:47}) as $\mathcal{D}$ varies as $(1+z_{s}^2)$. As we know from BJJ analysis that, when $\Lambda_{0}<1$ the dynamics of the population imbalance shows Josephson oscillation. However when the critical value of $\Lambda_{0}$ is crossed, there are two types of MQST as observed depending on the time average of population imbalance such as $\langle z\rangle<|z_{s}|\neq 0$ and $\langle z\rangle>|z_{s}|\neq 0$. These two types of MQST are known as running phase and $\pi$-phase modes of MQST. We also know that the range, $1<\Lambda_{0}<2$ specifies the regime of $\pi$-phase mode MQST and $\Lambda_{0}>2$ refers to the running phase mode MQST which is similar to the zero-phase mode MQST.

For small dissipation coefficient ($\tilde{\gamma}=1$) at low temperature ($k_{B}T/\hbar\Omega=1/5$), one may assume that the dynamics of the dissipative BJJ follows the standard results of the BJJ and the Josephson Hamiltonian \cite{ch6_41c} gives negative energy for the zero-phase mode and positive energy for the $\pi$-phase mode \cite{ch6_41c}. Now, in the $\pi$-phase mode, the system initially has large energy; so increasing $\Lambda_{0}$ does not change the flow of the relative phase between two wells in the DW potential as a result of which $\mathcal{D}$ decreases. But at the critical $\Lambda_{0}$, the interaction energy becomes equal to the steady state Josephson Hamiltonian energy which implies that at the critical point there is no flow of the relative phase between the wells. Further increase of $\Lambda_{0}$, implies that many body interaction energy becomes large compared to the Josephson Hamiltonian energy so that the flow is reversed and as a result $\mathcal{D}$ increases with increase of $\Lambda_{0}$. This is also similar to the case of zero-phase mode because in zero-phase mode initial Josephson Hamiltonian energy is negative and with increase of positive $\Lambda_{0}$, $\mathcal{D}$ increases. Although the behaviour of $\mathcal{D}$ as a function of $\Lambda_{0}$ in zero-phase mode is not shown in the text (since $\mathcal{D}\propto C^2\gamma \propto (1+\Lambda_{0})^2$), it is apparent that $\mathcal{D}$ increases with increase of $\Lambda_{0}$.

\subsection{Quantum fluctuation spectra}
For the study of the quantum fluctuation of number and phase we first choose the region from Fig. \ref{fig:1a}(a) where $U/K\ll1$ to explore the effect of the dissipation on fluctuation spectra in the coherent and near coherent regime. Our primary focus lies on the interplay of interaction and dissipation to study in these regimes of 1D dissipative BJJ.

From Eqs (\ref{eqn:58}) and (\ref{eqn:59}), in the zero-phase mode, the analytical expression for the spectrum of number fluctuation becomes
\begin{eqnarray}
S_{z}(\Delta)=\frac{\frac{2\tilde{\gamma}}{N}\coth({\frac{\hbar\Omega}{2k_{B}T})[4+\Delta^2]}}{\left[\Delta^2-\omega_{J}^2+\frac{\tilde{\gamma}^2}{2}\right]^2+\left[\omega_{J}^2\tilde{\gamma}^2-\frac{\tilde{\gamma}^4}{4} \right]}
\end{eqnarray}
and the spectrum of phase fluctuation can be written as
\begin{eqnarray}
S_{\theta}(\Delta)=\frac{\frac{2\tilde{\gamma}}{N}\coth({\frac{\hbar\Omega}{2k_{B}T})[\omega_{J}^2(1+\Lambda_{0})+\tilde{\gamma}^2+\Delta^2]}}{\left[\Delta^2-\omega_{J}^2+\frac{\tilde{\gamma}^2}{2}\right]^2+\left[\omega_{J}^2\tilde{\gamma}^2-\frac{\tilde{\gamma}^4}{4} \right]}
\end{eqnarray}
where $\tilde{\gamma}=\gamma N$, and Josephson frequency $\omega_{J}=\frac{2K}{\hbar}\sqrt{1+\Lambda_{0}}$. In the $\pi$-phase mode Josephson oscillation and the $\pi$-phase mode self-trapping regime the analytical expression for the spectrum of number fluctuation becomes
\begin{eqnarray}
S_{z}(\Delta)=\frac{\frac{2\tilde{\gamma}}{N}\coth({\frac{\hbar\Omega}{2k_{B}T})[4+\Delta^2]}}{\left[\Delta^2-\omega_{\pi}^2+\frac{\tilde{\gamma}^2}{2}\right]^2+\left[\omega_{\pi}^2\tilde{\gamma}^2-\frac{\tilde{\gamma}^4}{4} \right]}
\end{eqnarray}

\begin{eqnarray}
S_{z}(\Delta)=\frac{\frac{2\tilde{\gamma}}{N}\coth({\frac{\hbar\Omega}{2k_{B}T})[\frac{4}{\Lambda_{0}^2}+2\Delta^2-\frac{\Delta^2}{\Lambda_{0}^2}]}}{\left[\Delta^2-\omega_{\rm ST}^2+\frac{\tilde{\gamma}^2}{2}\right]^2+\left[\omega_{\rm ST}^2\tilde{\gamma}^2-\frac{\tilde{\gamma}^4}{4} \right]}
\end{eqnarray}
respectively and the corresponding spectrum of phase fluctuation are given by 
\begin{eqnarray}
S_{\theta}(\Delta)=\frac{\frac{2\tilde{\gamma}}{N}\coth({\frac{\hbar\Omega}{2k_{B}T})[\omega_{\pi}^2(1-\Lambda_{0})+\tilde{\gamma}^2+\Delta^2]}}{\left[\Delta^2-\omega_{\pi}^2+\frac{\tilde{\gamma}^2}{2}\right]^2+\left[\omega_{\pi}^2\tilde{\gamma}^2-\frac{\tilde{\gamma}^4}{4} \right]}
\end{eqnarray}

\begin{eqnarray}
S_{\theta}(\Delta)=\frac{\frac{2\tilde{\gamma}}{N}\coth({\frac{\hbar\Omega}{2k_{B}T})[\omega_{\rm ST}^2(\Lambda_{0}^2-1)(2\Lambda_{0}^2-1)+\tilde{\gamma}^2+\Delta^2]}}{\left[\Delta^2-\omega_{\rm ST}^2+\frac{\tilde{\gamma}^2}{2}\right]^2+\left[\omega_{\rm ST}^2\tilde{\gamma}^2-\frac{\tilde{\gamma}^4}{4} \right]}\nonumber
\end{eqnarray}
where $\omega_{\pi}=\frac{2K}{\hbar}\sqrt{1-\Lambda_{0}}$ and $\omega_{\rm ST}=\frac{2K}{\hbar}\sqrt{\Lambda_{0}^2-1}$ signifies the Josephson frequency and self-trapping frequency in $\pi$-phase mode of standard BJJ.

From the above expressions for the number and phase fluctuation spectra, we observe the following

1. When the detuning $\Delta=0$, both number and phase fluctuations persist $(S_{z}, S_{\theta}\neq 0)$ in zero and $\pi$-phase modes of 1D dissipative BJJ. Quantum fluctuations remain even at absolute zero temperature.

2. It is also clear that the term defining the peak position depends on the Josephson frequency $\omega_{J}$ and $\omega_{\pi}$ for zero and $\pi$-phase modes and $\omega_{\rm ST}$ for $\pi$-phase mode self-trapping regime with dissipation parameter $\tilde{\gamma}$. The contribution of $\tilde{\gamma}$ arises due to the induced bath effect on the system.  

\begin{figure}[h]
\centering
\includegraphics[width=8.0cm]{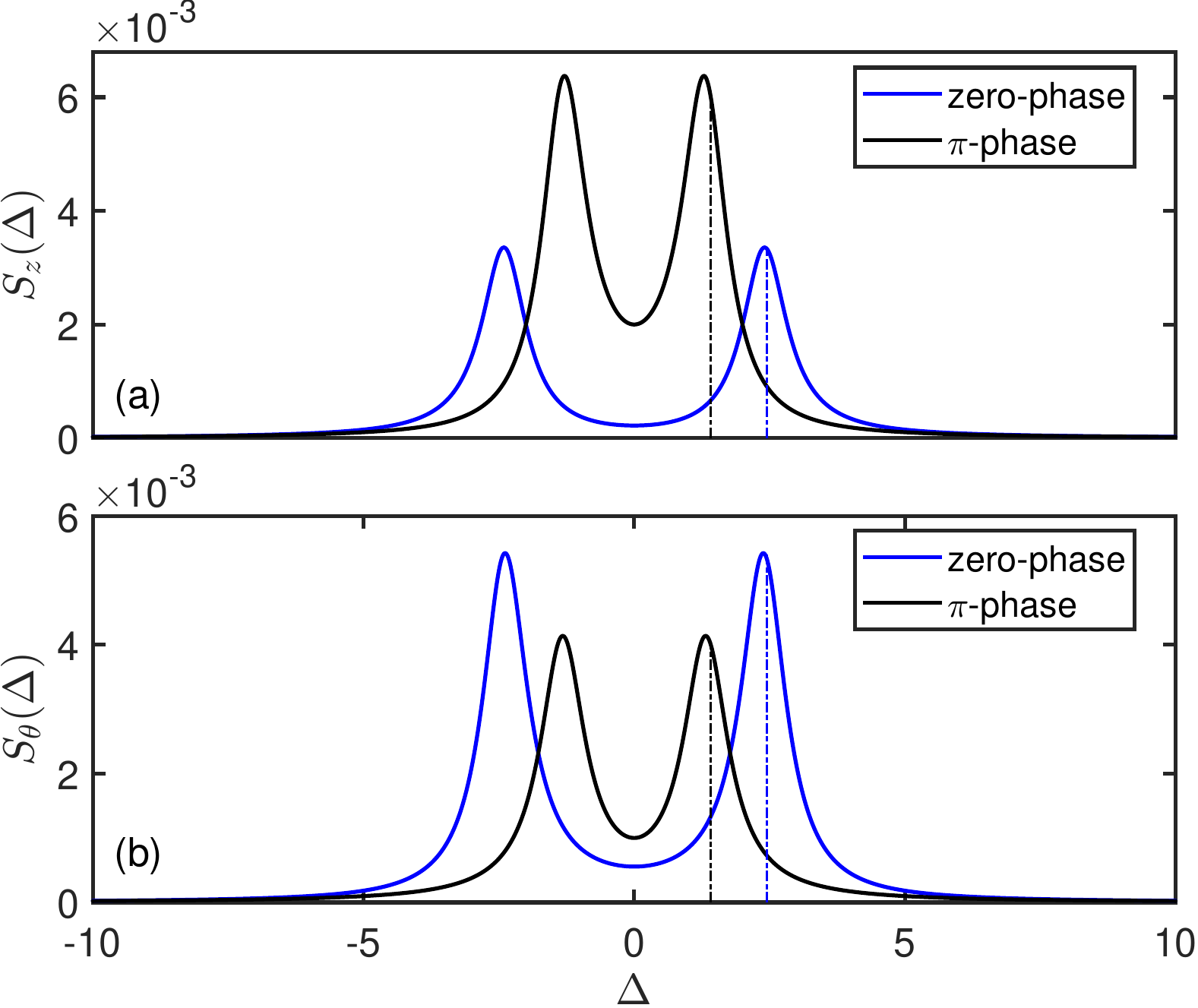}
\caption{(Color online) a) Variation of number fluctuation spectrum $\mathcal{S}_{z} (\Delta)$ as a function of detuning $\Delta$ for zero phase mode (solid blue line) and $\pi$-phase mode (solid black line). b) Variation of the phase fluctuation spectrum $\mathcal{S}_{\theta} (\Delta)$ as a function of the detuning $\Delta$ for zero phase mode (solid blue line) and $\pi$-phase mode (solid black line). Both the curves are plotted for small interaction limit $U/K=0.001$, small dissipation $\tilde{\gamma}=1$ limit with $N=1000$ and low temperature regime $k_{B}T/\hbar\Omega=1/10$. The blue and black dashed-dotted lines indicate the Josephson frequency for zero and $\pi$-phase mode.}
\label{fig:4a}
\end{figure}

For our numerical calculation, we consider first the low temperature limit $k_{B}T/\hbar\Omega=1/10$ and the weak interaction limit $U/K=0.001$. It is interesting to examine how the system behaves in the weak interaction regime in presence of weak dissipation. For that we choose $\tilde{\gamma}=1$ with total number of atoms $N=1000$.
In Fig. \ref{fig:4a}(a) and \ref{fig:4a}(b), we plot the number and phase fluctuation spectra as a function of detuning for weak interaction $(U/K=0.001)$ and weak dissipation limit with $\tilde{\gamma}=1$ in the low temperature regime for zero phase mode denoted by the solid blue line and the $\pi$-phase Josephson oscillation regime indicated by the solid black line. For each mode, we observe that two peaks appear where the dashed-dotted lines indicate the Josephson frequency in the respective mode. It is clear that exactly at the Josephson frequency the amplitude of the number and phase fluctuation spectra is maximum and at $\Delta=0$ both number fluctuation and phase fluctuation persist in the system for both zero and $\pi$-phase mode. However, we observe that in the $\pi$-phase mode the amplitude of the number fluctuation spectrum increases compared to the amplitude of the number fluctuation spectrum in the zero phase mode and the opposite behaviour is observed for the phase fluctuation spectra. We also observe that the depth which implies that the difference between the maximum and minimum of the amplitude of the spectra at $\Delta=0$ increases for the $\pi$-phase number fluctuation spectra compared to that for the zero phase. The reason behind this behaviour is that for a fixed many body interaction parameter the Josephson frequency for the zero phase mode is always greater than the Josephson frequency in $\pi$-phase mode. As a result when the Josephson frequency decreases the amplitude increases as well as the depth increases. However, we also note similar but opposite behaviour for the phase fluctuation spectra. It is also apparent that due to the presence of $\tilde{\gamma}$ in the analytical as well as numerical calculation of spectra the peak values of the number and phase fluctuation spectra are not exactly at the Josephson frequency in the $\pi$-phase mode of the standard BJJ. This also signifies that the peak frequency is dressed by $\tilde{\gamma}$. For both cases we observe that the number and phase fluctuation becomes close to zero for large detuning. This picture qualitatively describes the coherent behaviour of the 1D dissipative BJJ in the small interaction and the small dissipation limit where both the number and the phase of the atoms in the well oscillate with the Josephson frequency. 

\begin{figure}[h]
\centering
\includegraphics[width=8.0cm]{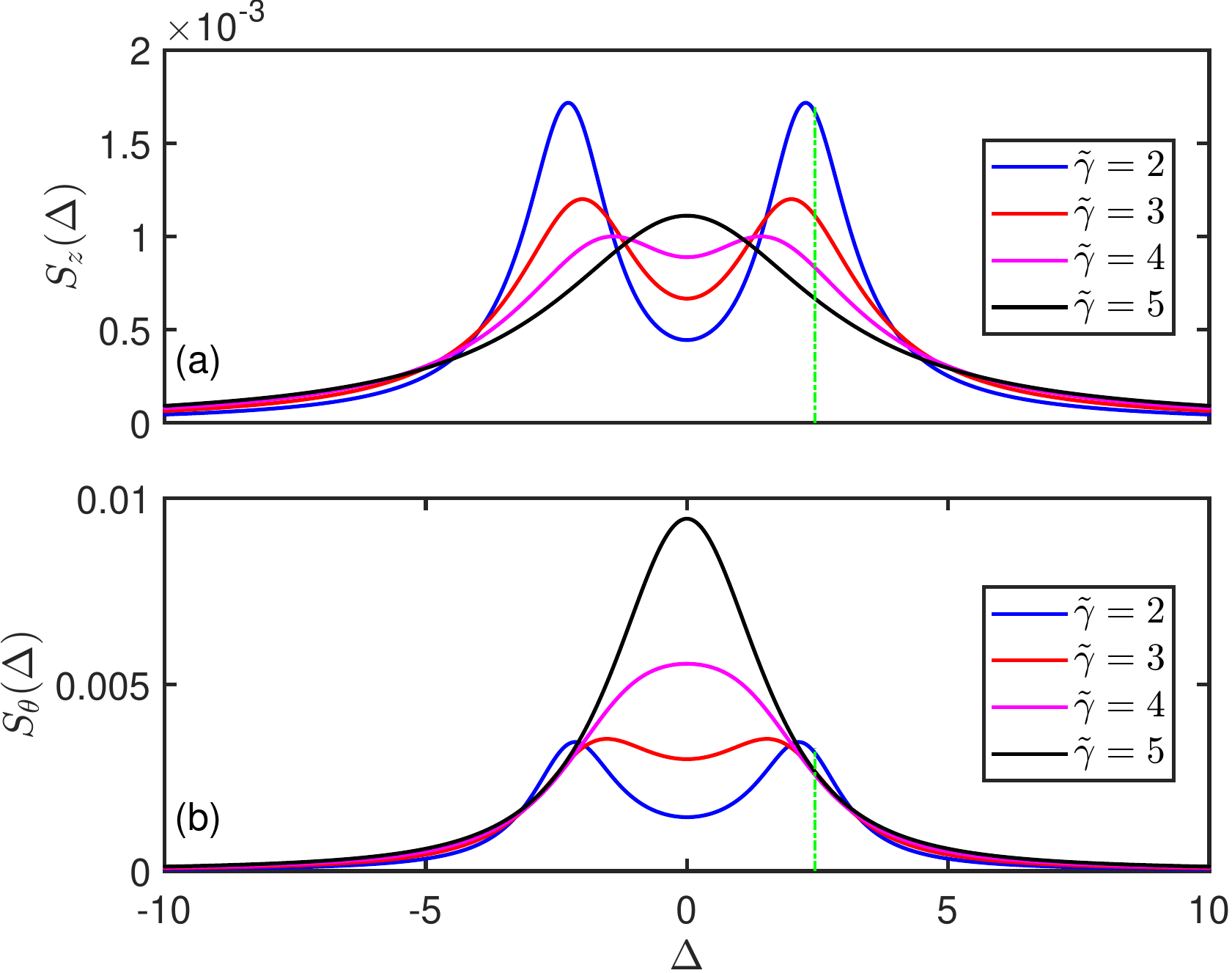}
\caption{(Color online) a) Variation of number fluctuation spectrum $\mathcal{S}_{z} (\Delta)$ as a function of detuning $\Delta$ with different dissipation coefficient $\tilde{\gamma}$ for zero phase mode. b) Variation of the phase fluctuation spectrum $\mathcal{S}_{\theta} (\Delta)$ as a function of the detuning $\Delta$ with different dissipation coefficient $\tilde{\gamma}$ for zero phase mode. Both the curves are plotted for small interaction limit $U/K=0.001$, with $N=1000$ and low temperature regime $k_{B}T/\hbar\Omega=1/10$. The green dashed-dotted line indicates the Josephson frequency $\omega_{J}$ in zero phase mode in absence of dissipation.}
\label{fig:5a}
\end{figure}

We now examine how this phase and number fluctuations change on increasing the dissipation in the system. In Fig. \ref{fig:5a}(a) and \ref{fig:5a}(b), we plot the number and phase fluctuation spectra as a function of detuning for weak interaction $(U/K=0.001)$ in the weak dissipation $(\tilde{\gamma}=2)$ to strong dissipation limit $(\tilde{\gamma}=5)$ in the low temperature regime for zero-phase mode. Here we also choose the total number of atoms to be $N=1000$. For $\tilde{\gamma}=2$ (solid blue line), we observe that the number and phase fluctuation spectra exhibit coherent behaviour. If we further increase $\tilde{\gamma} (=3$,  solid red line) the amplitude of the number fluctuation spectrum is reduced and also the depth decreases. We also note that the peak frequency is shifted away from the Josephson frequency which implies that the peak frequency is dressed by the $\omega_{J}$ and $\tilde{\gamma}$. Similar behaviour is observed in the phase fluctuation spectrum with a difference that the amplitude of the phase fluctuation spectrum slightly increases and also the peak frequency gets shifted slightly from the Josephson frequency. With further increase of $\tilde{\gamma}(=4$, solid magenta line), the amplitude of the number fluctuation spectrum is reduced and the depth at $\Delta=0$ is also reduced. However the two peaks originate in the number fluctuation spectrum which implies that the system still oscillates with a dressed Josephson frequency whereas the phase fluctuation spectrum shows a single peak originated at $\Delta=0$ and describes the phase fluctuation is in incoherent regime because the phase oscillates in each well between the ground and first excited state (of a single-particle state). In this case the tunneling of the particle is prohibited. For further increase of $\tilde{\gamma}(=5$, solid black line), both number and phase fluctuation spectra show the growth of a single peak at $\Delta=0$ which implies both number and phase of the atoms in the well oscillate with a frequency which is equal to the energy gap of the first excited state and the ground state (of a single particle state). So, by changing the dissipation coefficient in the low temperature regime with weak interaction one can realize a transition from coherent to incoherent regime. The coherent regime implies the standard BJJ picture where the system oscillates between each well with a frequency close to the Josephson frequency whereas for the incoherent regime the tunneling of the particles are prohibited.

\begin{figure}[h]
\centering
\includegraphics[width=8.0cm]{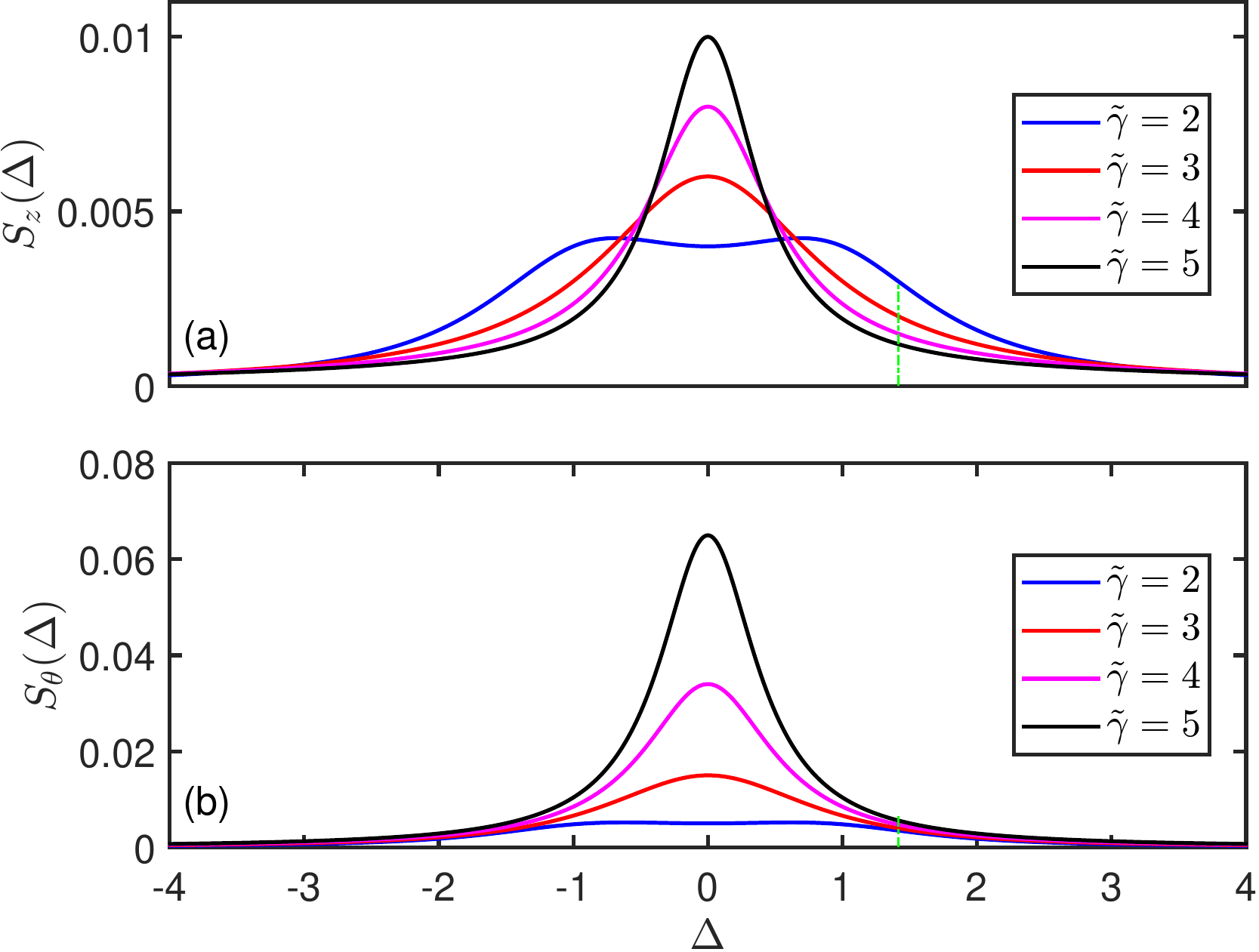}
\caption{(Color online) a) Variation of number fluctuation spectrum $\mathcal{S}_{z} (\Delta)$ as a function of detuning $\Delta$ with different dissipation coefficient $\tilde{\gamma}$ for $\pi$-phase Josephson oscillation regime. b) Variation of the phase fluctuation spectrum $\mathcal{S}_{\theta} (\Delta)$ as a function of the detuning $\Delta$ with different dissipation coefficient $\tilde{\gamma}$ for $\pi$-phase Josephson oscillation regime. Both the curves are plotted for small interaction limit $U/K=0.001$, with $N=1000$ and low temperature regime $k_{B}T/\hbar\Omega=1/10$. The green dashed-dotted line indicates the Josephson frequency $\omega_{\pi}$ in $\pi$-phase mode Josephson oscillation regime in absence of damping.}
\label{fig:5anew}
\end{figure}

We now examine the effect of dissipation in number and phase fluctuation spectra in the $\pi$-phase mode Josephson oscillation regime. In Fig. \ref{fig:5anew}(a) and \ref{fig:5anew}(b), we plot the number and phase fluctuation spectra as a function of detuning for weak interaction $(U/K=0.001)$ in the low temperature regime for several values of dissipation strength. We also choose that the total number of atom $N=1000$. For $\tilde{\gamma}=2$ (solid blue line), we see that the number and phase fluctuation spectra exhibit two peaks with small amplitude which are distinct from the $\pi$-phase mode Josephson frequency as represented by the green dashed-dotted line in Fig. \ref{fig:5anew}. With further increase of dissipation parameter $\tilde{\gamma}=3$, the number and the phase fluctuation spectra show a single peak at $\Delta=0$ which implies that the system is in the incoherent regime. With further increase of $\tilde{\gamma}=4$ and $\tilde{\gamma}=5$, the number and phase fluctuation spectra remain in the incoherent regime but the amplitude of the fluctuation spectra increases with increasing dissipation strength. Here almost for $\tilde{\gamma}=2$, the system makes a transition to the incoherent state whereas in the zero phase mode with $\tilde{\gamma}=4$ the system becomes incoherent in phase. We also observe that for large detuning both number and phase fluctuation become close to zero. In the $\pi$-phase mode with weak dissipation one can always realize the incoherent regime whereas in the zero phase mode one may observe a transition from coherent to incoherent regime for comparatively large dissipation. 

\begin{figure}[h]
\centering
\includegraphics[width=8.0cm]{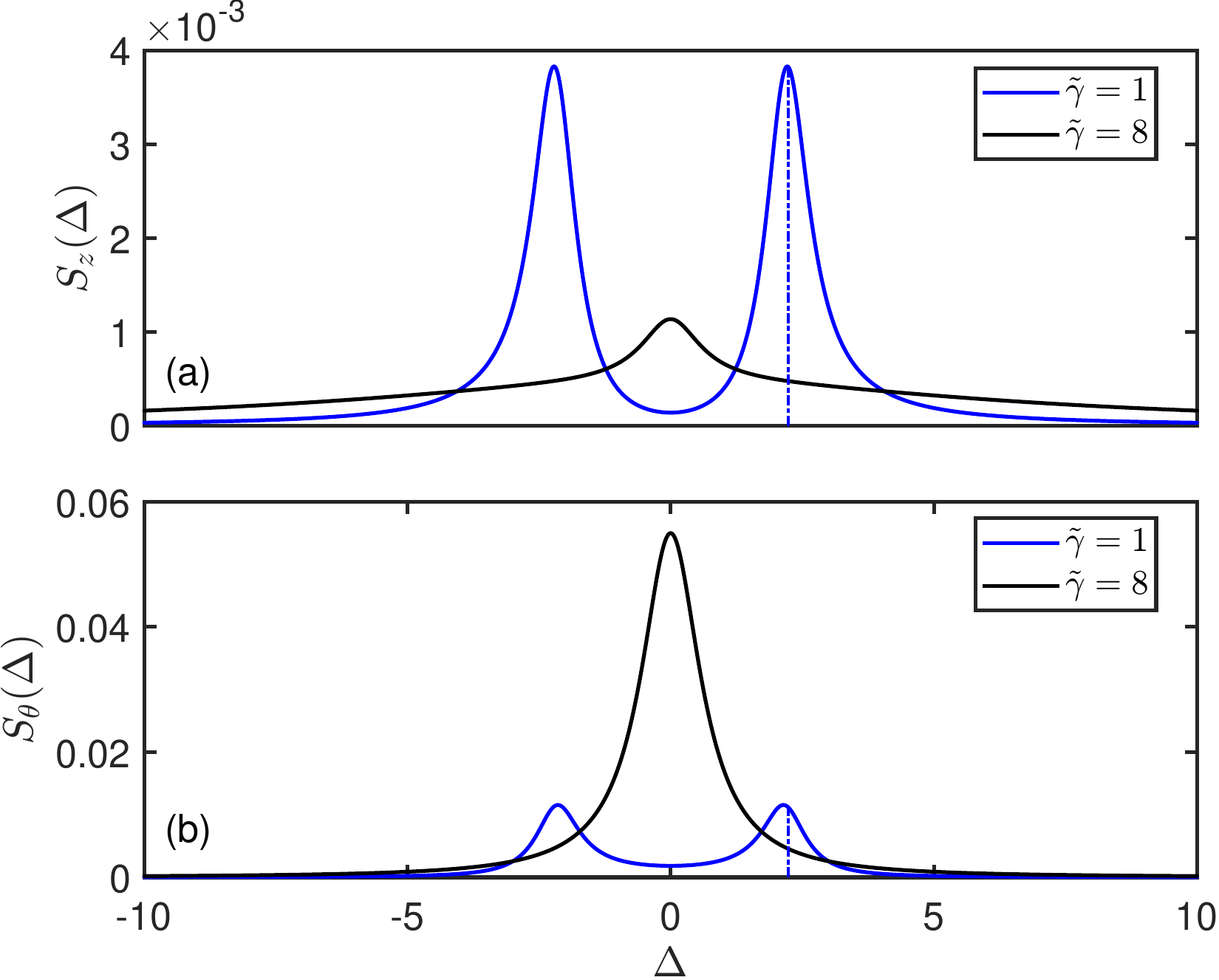}
\caption{(Color online) a) Variation of number fluctuation spectrum $\mathcal{S}_{z} (\Delta)$ as a function of detuning $\Delta$ for weak dissipation coefficient $\tilde{\gamma}=1$ (solid blue color) and strong dissipation coefficient $\tilde{\gamma}=8$ for $\pi$-phase mode self-trapping regime. b) Variation of the phase fluctuation spectrum $\mathcal{S}_{\theta} (\Delta)$ as a function of the detuning $\Delta$ for weak dissipation coefficient $\tilde{\gamma}=1$ (solid blue color) and strong dissipation coefficient $\tilde{\gamma}=8$ for $\pi$-phase mode self-trapping regime. Both the curves are plotted for interaction value $U/K=0.003$, with $N=1000$ and low temperature regime $k_{B}T/\hbar\Omega=1/10$. The blue dashed-dotted line indicates the self-trapping  frequency $\omega_{\rm ST}$ in $\pi$-phase mode.}
\label{fig:5anew1}
\end{figure}

In Fig. \ref{fig:5anew1}(a) and \ref{fig:5anew1}(b), we plot the number and phase fluctuation spectra as a function of detuning for interaction $U/K=0.003$ in the low temperature regime for weak dissipation $(\tilde{\gamma}=1)$ and strong dissipation limit $(\tilde{\gamma}=8)$ in $\pi$-phase self-trapping regime. For our numerical calculation, we choose total number of atoms $N=1000$. In the $\pi$-phase self-trapping regime, it is necessary to choose the value of interaction energy $U/K>0.002$ such that the many-body interaction parameter $\Lambda_{0}$ becomes greater than unity. We observe that for weak dissipation $(\tilde{\gamma}=1$ the solid blue line) the number and phase fluctuation spectra show the two peaks implying the coherent behaviour where the peak amplitude of the number and phase fluctuation spectra is at $\pi$-phase mode self-trapping frequency denoted by the dashed-dotted blue line in Fig. \ref{fig:5anew1}. This picture qualitatively describes the coherent behaviour of the 1D dissipative BJJ in the small dissipation limit where both number and phase of the atoms in the wells oscillate with non-zero average value and the frequency of the oscillation is determined by the $\pi$-phase mode self-trapping frequency. Now for strong dissipation $(\tilde{\gamma}=8)$, we observe that a single peak appear at $\Delta=0$ both in number and phase fluctuation spectra. However the amplitude of the phase fluctuation spectra is always greater than that for the number fluctuation spectra. This picture qualitatively describes the effect of dissipation and the transition from coherent to incoherent regime in the $\pi$-phase mode self-trapping regime of 1D dissipative BJJ.

\section{Conclusion}\label{sec:6}
In this paper, we have considered a nonlinear dissipative BJJ in presence of quantum noise. We have shown that the dissipative BJJ equations in presence of weak noise follow linear Langevin dynamics of relative phase. From this Langevin description of the relative phase, we derive the analytical formula for the phase diffusion coefficient for zero and $\pi$-phase modes of BJJ. We further derive the analytical formula of the number fluctuation spectra and the phase fluctuation spectra to analyze its dependence on interaction energy and dissipation. The main conclusions of this study can be summarized as follows:
\begin{itemize}
\item  We have formulated the dissipative BJJ equations (\ref{eqn:imbalance}) and (\ref{eqn:phase}) within the framework of a $c$-number description of noise within Born-Markov approximation, where the equilibrium distribution follows Wigner thermal canonical distribution. The equations are classical looking in form but quantum mechanical in content.

\item From the dissipative BJJ dynamics around the steady states, the phase diffusion coefficient and the spectra of number and phase fluctuation are derived.

\item Our numerical simulations in zero-phase mode suggest that $\mathcal{D}$ depends on the system interaction parameter $\Lambda_{0}$, dissipation $\gamma$ and temperature. We show that $\mathcal{D}$ increases with increase in temperature for small dissipation. However, if dissipation is comparatively large the variation of $\mathcal{D}$ with temperature slows down.

\item Our numerical simulations in $\pi$-phase modes for small dissipation and low temperature regime suggest that the phase diffusion coefficient $\mathcal{D}$ exhibits an interesting phase transition like behaviour between the Josephson oscillation and the MQST regime. In the Josephson oscillation regime $\mathcal{D}$ decreases with increase of $\Lambda_{0}$ and in the MQST regime above the critical $\Lambda_{0}$, $\mathcal{D}$ increases with increase of $\Lambda_{0}$. $\mathcal{D}$ does not show bifurcation behaviour although the steady state population imbalance shows the bifurcation behaviour. 

\item We have also analyzed the coherence factor of the dissipative BJJ as defined in equation (\ref{eqn:50new}) where the averaging is done using Wigner thermal canonical distribution. We analyze numerically the coherence factor as a function of $U/K$ and show that in the strong tunneling regime $U/K\ll1$, the system becomes coherent and for the opposite case $U/K\gg1$, the system goes into the incoherent state. We also analyze the coherent and incoherent behaviour of the system in terms of temperature for the fixed $U/K$ ratios where for low temperature the system remains in the coherent state and for high temperature the system goes into a incoherent state. In the $\pi$-phase mode self-trapping regime, the coherence factor suddenly falls from unity to zero with small change in $U/K$ ratio but it exhibits similar temperature dependence as in zero-phase mode. Our analysis of coherence factor reveals that the BJJ has a higher degree of coherence which is in good agreement with the experimental observation \cite{ch6_39}.

\item We have also numerically analyzed the coherent and incoherent behaviour of the system in terms of quantum fluctuation spectra of population imbalance and phase difference in presence of dissipation. We observe that in the weak dissipation limit the fluctuation spectra show two peak at the Josephson frequency whereas for the large dissipation limit the fluctuation spectra show a single peak which signifies the incoherent state of the system. This is reminiscent of Autler-Townes effect in quantum optics.
\end{itemize}

Although the calculations in this work are specific to a scalar BECs, but the concepts and ideas presented in this work are also applicable to the two-component BEC where the dynamics of the phase diffusion coefficient can be studied in terms of internal BJJ in presence of noise and dissipation. It would be interesting to investigate how phase diffusion coefficient behave with change of the population transfer between the modes of the two-component BEC.

\end{document}